\documentclass[aps, prl, reprint, superscriptaddress]{revtex4-1}

\usepackage[utf8]{inputenc}
\usepackage{amsmath}
\usepackage{amssymb}
\usepackage{hyperref}
\usepackage{graphicx}
\usepackage{dcolumn}
\usepackage{bm}
\usepackage{soul}
\usepackage{xspace}
\usepackage{verbatim}
\usepackage{color}

\newcommand{\ket}[1]{\left| #1 \right\rangle}

\hypersetup{colorlinks=true,linkcolor=blue,citecolor=blue, filecolor=blue,urlcolor=blue,breaklinks=true}

\begin{document}

\title{Multi-Level Variational Spectroscopy using a Programmable Quantum Simulator}

\newcommand{\SIQSE}{\affiliation{1}{Shenzhen Institute for Quantum Science and Engineering, Southern University of Science and Technology, Shenzhen, Guangdong, China}}
\newcommand{\IQA}{\affiliation{2}{International Quantum Academy, Shenzhen, Guangdong, China}}
\newcommand{\GDKL}{\affiliation{3}{Guangdong Provincial Key Laboratory of Quantum Science and Engineering, Southern University of Science and Technology, Shenzhen, Guangdong, China}}
\newcommand{\DPHY}{\affiliation{4}{Department of Physics, Southern University of Science and Technology, Shenzhen, Guangdong, China}}
\newcommand{\HFNL}{\affiliation{5}{Shenzhen Branch, Hefei National Laboratory, Shenzhen, China}}
\newcommand{\UESTC}{\affiliation{6}{Institute of Fundamental and Frontier Sciences,
University of Electronic Science and Technology of China, Chengdu, China}}

\author{Zhikun Han}
\thanks{These authors have contributed equally to this work.}
\affiliation{\SIQSE}\affiliation{\IQA}\affiliation{\GDKL}

\author{Chufan Lyu}
\thanks{These authors have contributed equally to this work.}
\affiliation{\UESTC}

\author{Yuxuan Zhou}
\thanks{These authors have contributed equally to this work.}
\affiliation{\SIQSE}\affiliation{\IQA}\affiliation{\GDKL}\affiliation{\DPHY}

\author{Jiahao Yuan}
\affiliation{\SIQSE}\affiliation{\IQA}\affiliation{\GDKL}\affiliation{\DPHY}
\author{Ji Chu}
\affiliation{\SIQSE}\affiliation{\IQA}\affiliation{\GDKL}
\author{Wuerkaixi Nuerbolati}
\affiliation{\SIQSE}\affiliation{\IQA}\affiliation{\GDKL}
\author{Hao Jia}
\affiliation{\SIQSE}\affiliation{\IQA}\affiliation{\GDKL}
\author{Lifu Nie}
\affiliation{\SIQSE}\affiliation{\IQA}\affiliation{\GDKL}
\author{Weiwei Wei}
\affiliation{\SIQSE}\affiliation{\IQA}\affiliation{\GDKL}
\author{Zusheng Yang}
\affiliation{\SIQSE}\affiliation{\IQA}\affiliation{\GDKL}
\author{Libo Zhang}
\affiliation{\SIQSE}\affiliation{\IQA}\affiliation{\GDKL}
\author{Ziyan Zhang}
\affiliation{\SIQSE}\affiliation{\IQA}\affiliation{\GDKL}

\author{Chang-Kang Hu}
\affiliation{\SIQSE}\affiliation{\IQA}\affiliation{\GDKL}
\author{Ling Hu}
\affiliation{\SIQSE}\affiliation{\IQA}\affiliation{\GDKL}
\author{Jian Li}
\affiliation{\SIQSE}\affiliation{\IQA}\affiliation{\GDKL}
\author{Dian Tan}
\affiliation{\SIQSE}\affiliation{\IQA}\affiliation{\GDKL}

\author{Abolfazl Bayat}
\email{abolfazl.bayat@uestc.edu.cn}
\affiliation{\UESTC}

\author{Song Liu}
\email{lius3@sustech.edu.cn}
\affiliation{\SIQSE}\affiliation{\IQA}\affiliation{\GDKL}\

\author{Fei Yan}
\email{yanfei@baqis.ac.cn}
\altaffiliation[Present address: ]{Beijing Academy of Quantum Information Sciences, Beijing, China}
\affiliation{\SIQSE}\affiliation{\IQA}\affiliation{\GDKL}

\author{Dapeng Yu}
\email{yudp@sustech.edu.cn}
\affiliation{\SIQSE}\affiliation{\IQA}\affiliation{\GDKL}\affiliation{\DPHY}

\begin{abstract}
\textbf{Energy spectroscopy is a powerful tool with diverse applications across various disciplines. The advent of programmable digital quantum simulators opens new possibilities for conducting spectroscopy on various models using a single device. Variational quantum-classical algorithms have emerged as a promising approach for achieving such tasks on near-term quantum simulators, despite facing significant quantum and classical resource overheads. Here, we experimentally demonstrate multi-level variational spectroscopy for fundamental many-body Hamiltonians using a superconducting programmable digital quantum simulator. By exploiting symmetries, we effectively reduce circuit depth and optimization parameters allowing us to go beyond the ground state. Combined with the subspace search method, we achieve full spectroscopy for a 4-qubit Heisenberg spin chain, yielding an average deviation of 0.13 between experimental and theoretical energies, assuming unity coupling strength. Our method, when extended to 8-qubit Heisenberg and transverse-field Ising Hamiltonians, successfully determines the three lowest energy levels. In achieving the above, we introduce a circuit-agnostic waveform compilation method that enhances the robustness of our simulator against signal crosstalk. Our study highlights symmetry-assisted resource efficiency in variational quantum algorithms and lays the foundation for practical spectroscopy on near-term quantum simulators, with potential applications in quantum chemistry and condensed matter physics.}
\end{abstract}

\maketitle

Energy spectrum is a unique fingerprint of matter which makes spectroscopy~\cite{jones2006ultracold,fischer2007scanning,basov2014colloquium,ulbricht2011carrier} an indispensable tool in a range of scientific fields, including exoplanet composition determination, climate monitoring, and material synthesis. 
Recent advances in quantum simulation have led to the emergence of analog and digital quantum simulators implemented on various physical platforms, such as cold atoms~\cite{bloch2022superfluid}, ion-traps~\cite{lanyon2011universal,zhang2017observation}, nuclear magnetic resonance systems~\cite{li2017measuring}, photonic chips~\cite{wang2017experimental,zhong2020quantum,carolan2020variational}, Rydberg atoms~\cite{keesling2019quantum}, and superconducting circuits~\cite{salathe2015digital,wang2020efficient,karamlou2021analyzing,neill2021accurately,han2021experimental,braumuller2022probing,Zhang2022Digital,shi2022observing}. Although analog quantum simulators have been employed for spectroscopy by means of Fourier transform in a single-excitation manifold~\cite{roushan2017spectroscopic}, their scalability and applicability to arbitrary Hamiltonians remain limited.

Digital quantum simulators, based on programmable quantum circuits, provide remarkable versatility and enable the study of arbitrary Hamiltonians~\cite{mcardle2020quantum}. Variational quantum algorithms, such as the Variational Quantum Eigensolver (VQE)~\cite{peruzzo2014variational}, present a promising avenue for exploring practical quantum applications using near-term quantum simulators. By employing a hybrid approach that combines a quantum simulator with a classical optimizer, VQE simulations have successfully tackled ground state problems of many-body systems in quantum chemistry~\cite{google2020hartree,o2016scalable,kandala2017hardware,hempel2018quantum,colless2018computation,nam2020ground} and condensed matter physics~\cite{kokail2019self,sagastizabal2021variational}. However, the considerable resource overheads associated with both quantum and classical components have restricted VQE spectroscopy to small system sizes, predominantly targeting only two eigenstates~\cite{Tilly2020IBM,gocho2023excited,santagati2018witnessing}, with the exception of Ref.~\cite{colless2018computation} which approximates four eigenenergies in a two-qubit system.
In order to make quantum variational spectroscopy extensible, it is crucial to address resource overheads by simplifying the ansatz and reducing variational parameters. In fact, symmetry is a fundamental property that is found in every building block of our universe. The interaction between particles in strongly correlated many-body systems reveals several forms of symmetries. Theoretical proposals have suggested utilizing symmetries as a potential approach~\cite{meyer2023exploiting,Lyu2023symmetryenhanced,bravyi2020obstacles,gard2020efficient}. Nonetheless, conclusive experimental evidence remains to be provided.

In this article, we present an experimental demonstration of multi-level energy spectroscopy for fundamental many-body Hamiltonians using VQE on a superconducting digital quantum simulator. By employing a mix of symmetry-preserving ansatzes, we significantly reduce circuit depth and the number of variational parameters while targeting multiple energy levels in conjunction with the Subspace Search VQE (SSVQE) method~\cite{nakanishi2019subspace}. 
We also introduce a circuit compiling method that conveniently corrects for both spatial and temporal signal crosstalk, improving the consistency of the gate performance in our device.
We achieve full variational quantum spectroscopy for all 16 states of a 4-qubit Heisenberg spin chain, with the measured energies deviating from their theoretical values by $0.13$ on average, assuming a unity coupling strength.
We further extend our approach to 8-qubit Heisenberg and transverse-field Ising Hamiltonians, extracting the three lowest energy eigenstates and demonstrating the potential of our method for multi-level spectroscopy in many-body systems. 
Intriguingly, we find that in the cases where the cost function of the VQE algorithm contains several terms, the best convergence result is obtained using a hybrid optimizer that combines the gradient-free Nelder-Mead method with the gradient-based Adam method.

\begin{figure*}
  \centering
  \includegraphics[width=0.85\textwidth]{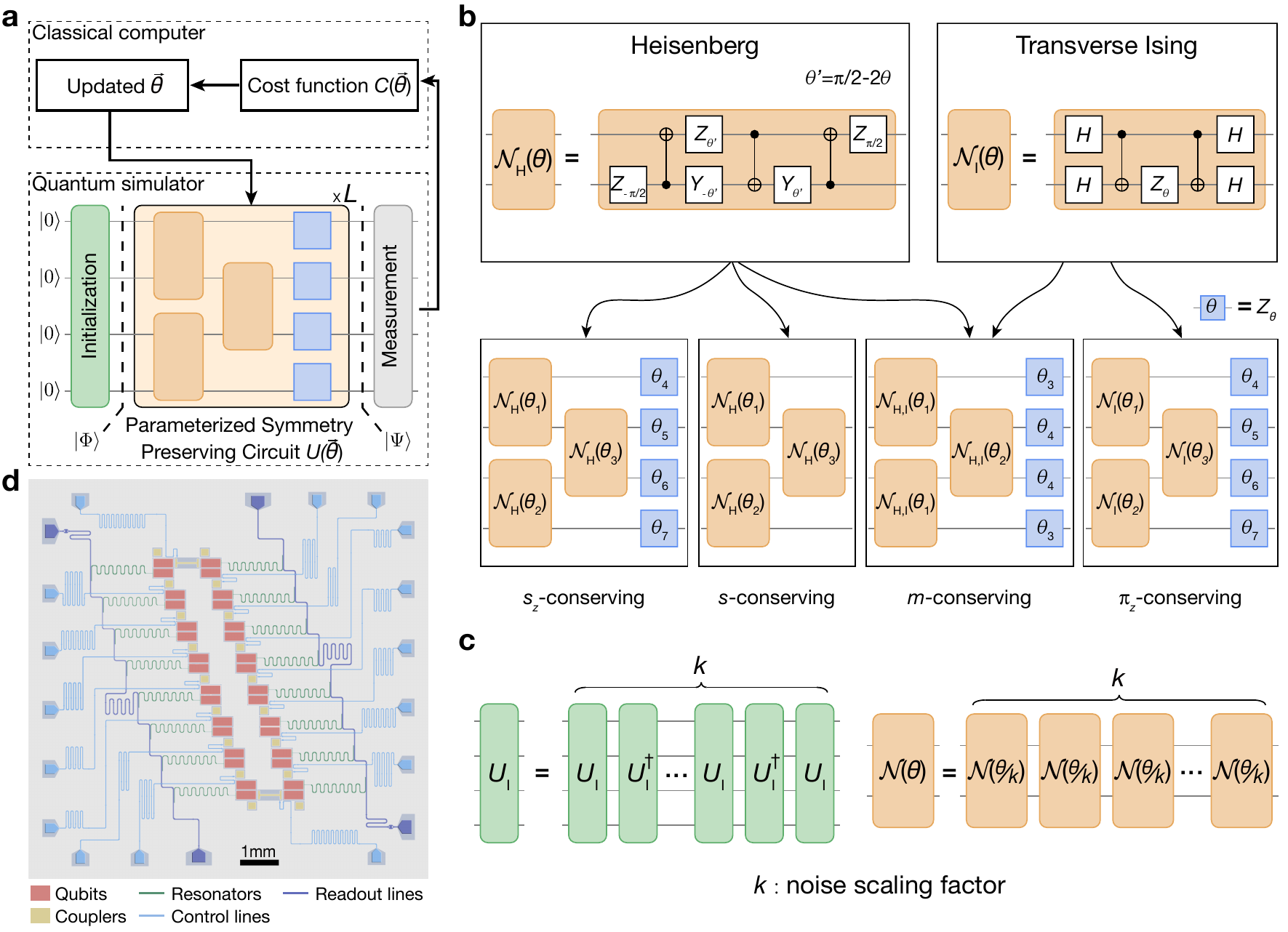}
  \caption[]
  {\textbf{Variational quantum algorithm with symmetry-preserving circuits.}
  \textbf{a,} Schematic diagram of a variational quantum algorithm which is realized on a combination of a quantum simulator and a classical computer. The quantum circuit includes an initialization circuit $U_I$, a parameterized symmetry-preserving circuit $U(\vec{\theta})$, and a measurement unit. The measurement results are fed into a classical optimizer for minimizing a cost function $C(\vec{\theta})$ through iteratively updating the parameters $\vec{\theta}$.
  \textbf{b,} Ansatz circuits for preserving different symmetries. In the top boxes are the two-qubit modular circuits used for the Heisenberg and transverse-field Ising problems. In the bottom are ansatz circuits for preserving $z$-polarization ($s_z$), total spin ($s$), mirror symmetry ($m$), and $zz$-product ($\pi_z$), respectively.
  \textbf{c,} Error mitigation using zero-noise extrapolation. $U_{\mathrm{I}}$ and $U_{\mathrm{I}}^\dagger$ are the initialization circuit and its reversal. Note that the $\mathcal{N}(\theta/k)$ circuit is identical to the original $\mathcal{N}(\theta)$ circuit except for single-qubit rotations according to the decomposition in (\textbf{b}). 
  \textbf{d,} False-colour micrograph of the device. }
  \label{Fig1_schematic}
\end{figure*}

As illustrated in Fig.~\ref{Fig1_schematic}\textbf{a}, the VQE algorithm contains a quantum hardware unit that executes a quantum circuit consisting of an initialization part, a multi-layer parameterized circuit, and measurement. The initialization circuit prepares an initial state of a $N$-qubit system through unitary operation $|\Phi\rangle{=}U_{\mathrm{I}}|0\rangle^{\otimes{N}}$. The parameterized circuit, however, is described by a unitary operation $U(\vec{\theta})$, where $\vec{\theta}=(\theta_1, \theta_2, \dots, \theta_l)$ are variational parameters. The action of $U(\vec{\theta})$ on the initial state creates the output state $|\Psi(\vec{\theta})\rangle{=}U(\vec{\theta})|\Phi\rangle$ for which an observable can be efficiently measured. A proper cost function $C(\vec{\theta})$ is calculated from the measured data which is then fed into a classical optimizer to be minimized through updating the circuit parameters $\vec{\theta}$, a training process until an optimal set of parameters $\vec{\theta}^*$ is obtained.

In comparison to typical VQE algorithms which find the ground state of a given Hamiltonian with the cost function being its expectation value~\cite{peruzzo2014variational,kokail2019self,google2020hartree,kandala2017hardware,o2016scalable,hempel2018quantum,lyu2022variational}, we employ a combination of several techniques to extend the spectrum and target more excited states.
For a system with known symmetries, we can prepare an initial state with a specific set of symmetry numbers and then apply the corresponding symmetry-preserving circuit such that the search is restricted within a certain manifold.
In this way, one can easily obtain the lowest energy eigenstate in that manifold -- but an excited state to the whole system. 
For cases when the states are indistinguishable by symmetry or when it is hard to prepare the initial state, simulating the excited states requires either adding additional penalizing terms in the cost function~\cite{Higgott2019variationalquantum,santagati2018witnessing} or simultaneous learning of multiple eigenstates through special training methods, such as SSVQE~\cite{nakanishi2019subspace}.
In our experiment, we use a general form for the cost function 
\begin{equation}
  C(\vec{\theta}) = \sum_{i=1}^P w_i \langle \Phi_i | U^{\dagger}(\vec{\theta}) H U(\vec{\theta}) | \Phi_i \rangle  \;,
  \label{eq:SSVQE_cost}
\end{equation}
where ${\{ \ket{\Phi_i} \}}_{i=1}^P$ are a set of $P$ orthogonal initial states and $w_i$ (real positive) are their weights with condition $w_1 {>} w_2 {>} \cdots {>} w_P$. Note that the output states $\ket{\Psi_i} {=} U(\vec{\theta}) \ket{\Phi_i}$ are naturally orthogonal due to the unitarity of $U(\vec{\theta})$. 
With proper constraints imposed on these weights, one obtains $\ket{E_i}{\approx}|\Psi_i(\vec{\theta}^*)\rangle$ through classical minimization of the cost function, provided that the parameterized circuit exhibits adequate expressibility. 

To understand how symmetry is integrated into the algorithm, let us first examine the well-known Heisenberg spin chain model that contains $N$ spin-1/2 particles. The Hamiltonian is expressed by
\begin{equation}
  H = J \sum_{i=1}^{N-1} \boldsymbol{\sigma}^{i} \cdot \boldsymbol{\sigma}^{i+1}  \;,
  \label{eq:Heis_Ham}
\end{equation}
where $J{>}0$ is the coupling strength and $\boldsymbol{\sigma}^{i}{=}(\sigma_x^i, \sigma_y^i, \sigma_z^i)$ is the vector of Pauli operators at site $i$. The Heisenberg Hamiltonian in Eq.~\eqref{eq:Heis_Ham} supports several symmetries including: (i) the magnetization in every direction, i.e.\ $[H, S_\alpha]{=}0$, where $S_\alpha{=}\frac{1}{2} \sum_i \sigma_\alpha^i$ with $\alpha{=}x,y,z$; (ii) the total spin,
i.e.\ $[H, S_\mathrm{tot}^2]{=}0$, where $S_\mathrm{tot}^2 {=} S_x^2 {+} S_y^2 {+} S_z^2$; (iii) mirror symmetry, i.e.\ $[H, \mathcal{M}]{=}0$, where the mirror operator $\mathcal{M}$ is defined as $\mathcal{M}\ket{b_1 b_2 \dots b_N}{=}\ket{b_N \dots b_2 b_1}$ with $\ket{b_i}{=}\ket{0}$ or $\ket{1}$ is the qubit state at site $i$. These symmetries imply that all the eigenstates $\ket{E_j}$ have a specific total spin $s$, namely $S_\mathrm{tot}^2 \ket{E_j} {=} s(s+1)\ket{E_j}$, $z$-magnetization $s_z$, given by $S_z \ket{E_j} {=} s_z \ket{E_j}$ in which  $-s \leq s_z \leq s$, and the mirror symmetry $m{=}{\pm} 1$, given by $\mathcal{M}\ket{E_j}{=} m \ket{E_j}$.

In general, each layer of our parameterized circuit contains two consecutive sets of two-qubit operations $\mathcal{N}(\theta)$, acting on odd and even bonds, followed by a series of single-qubit rotations (see Fig.~\ref{Fig1_schematic}\textbf{a}).
In the case of Heisenberg model, $\mathcal{N}(\theta)$ acting on qubits $i$ and $i{+}1$ is chosen to be
\begin{equation}
  \mathcal{N}_{\mathrm{H}}(\theta)=e^{i\theta \left(\sigma_x^i \sigma_x^{i+1}+ \sigma_y^i \sigma_y^{i+1}+ \sigma_z^i \sigma_z^{i+1}\right)}   \;.
\end{equation}
An explicit design of $\mathcal{N}_{\mathrm{H}}(\theta)$ as well as symmetry-preserving circuits are shown in Fig.~\ref{Fig1_schematic}\textbf{b}.
Since $[\mathcal{N}_{\mathrm{H}}(\theta),S_\mathrm{tot}^2]{=}0$, to conserve the total spin $s$ no single-qubit rotation is needed.
To conserve $s_z$, the single-qubit operations are chosen to be $z$-rotations, namely $Z_\theta=e^{-i\theta\sigma_z /2}$.
Note that the $s$-conserving circuit readily conserves $s_z$.
The mirror symmetry can be easily preserved by choosing the gate parameters symmetrically with respect to the middle of the chain, and it may be used in combination with the other symmetries.

To diminish the effect of noise and improve the accuracy of the extracted energies, we employ the Zero-Noise Extrapolation (ZNE) method~\cite{endo2018practical,kandala2019error,kim2023scalable} for error mitigation. To perform ZNE, we rescale the circuit depths by a factor of $k$ while maintaining the same unitary. 
As shown in Fig.~\ref{Fig1_schematic}\textbf{c}, for the parameterized circuit $U(\vec{\theta})$, every $\mathcal{N}(\theta)$ operation is divided into $k$ consecutive repetitions of $\mathcal{N}(\theta/k)$. Such a treatment allows the noise scaling factor $k$ to be any positive integer.
For the initialization circuit $U_{\mathrm{I}}$, we append the $U_{\mathrm{I}} U_{\mathrm{I}}^\dagger$ pairs, which means $k$ can only be odd-numbered.
To access more choices in $k$ and improve the statistics, for an even-numbered $k$, we perform both cases of $k{=}2n{-}1$ and $k{=}2n{+}1$ for the initialization circuit and take the averaged outcome.

\begin{figure*}
  \centering
  \includegraphics[width=0.85\textwidth]{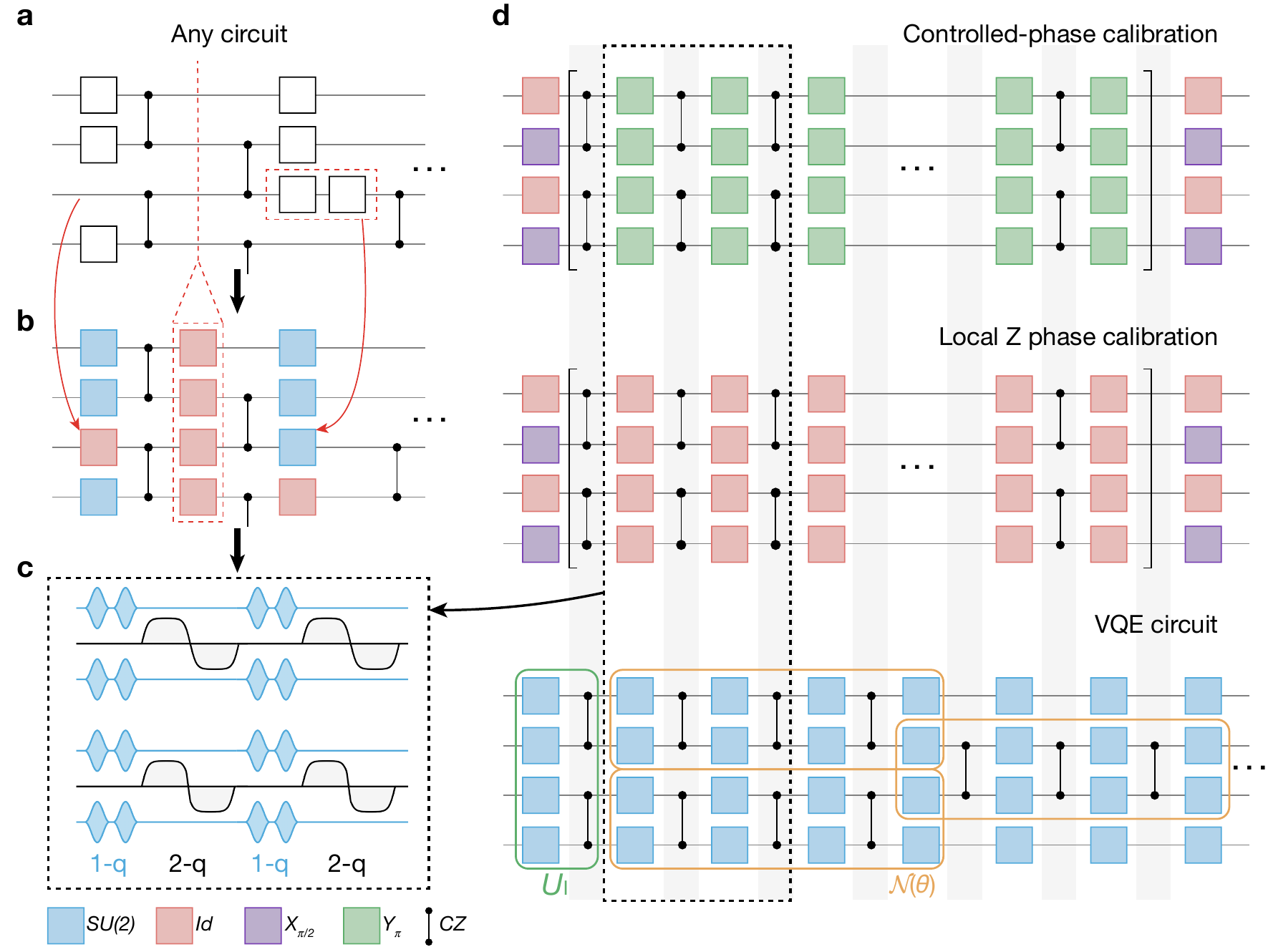}
  \caption[]
  {\textbf{Circuit compilation using the isomorphous waveform method.}
  \textbf{a-c,} Step-by-step compilation. An arbitrary quantum circuit composed of modular single- and two-qubit gates (\textbf{a}) can be rearranged into periodic single-qubit (1-q) and two-qubit (2-q) gate cycles (\textbf{b}). Red arrows indicate the identity filling and single-qubit gate squeeze operations. All SU(2) operations including identity operations are subsequently compiled into two $\pi/2$ pulses according to the U3 decomposition (\textbf{c}). 
  \textbf{d,} Gate sequences used in the controlled-phase calibration, the local Z phase calibration, and the VQE experiment for the Heisenberg model in which the initialization circuit (red solid box) and the $\mathcal{N}(\theta)$ circuit (brown solid box) are indicated. The black dashed box indicates a section of the three circuits that share a common waveform as illustrated in (\textbf{c}).}
  \label{Fig2_circuit_complilation}
\end{figure*}

Our experiment is performed on a superconducting quantum processor with a ring of 16 transmon qubits (Fig.~\ref{Fig1_schematic}\textbf{d}) mounted inside a dilution refrigerator. A tunable coupler - also a transmon qubit - connects each pair of neighboring qubits and serves as an independent control knob for adjusting their coupling strength. A shared control line is used for delivering both the single-qubit gate signal to the qubit and the two-qubit gate signal to an adjacent coupler, reducing the wiring efforts both on the chip and inside the refrigerator. More details about the device can be found in the Methods section and its 8-qubit predecessor \cite{chu2023scalable}. In this experiment, we use the coupler-assisted adiabatic controlled-$Z$ (CZ) gate~\cite{xu2020high} in conjunction with the net-zero pulsing technique to suppress dephasing from low-frequency flux noise \cite{rol2019fast}. The CZ gate fidelities when simultaneously applied are nearly 99\% on average.

The VQE algorithms and the ZNE error mitigation often involve long circuits with parallel gate operations. In practice, it is commonly seen that the performance of modular single- and two-qubit gates is inconsistent between calibration and final implementation, as the control signals are distorted due to the spatial and temporal crosstalk effect that varies with circuits. To combat this, we implement a generally applicable compilation strategy named \textit{isomorphic waveform}. 
The key idea is to maximally match the waveforms used during calibration and subsequent implementations by rearranging an arbitrary quantum circuit into interleaved layers of single- and two-qubit gates and translating them into well-timed pulses (Figs.~\ref{Fig2_circuit_complilation}\textbf{a}-\textbf{c}). 
First, for two-qubit gates that cannot be parallelized, one can always split them and insert a layer of identity gates between them (identity filling). 
Then, multiple concatenated single-qubit gates are squeezed into a single SU(2) operation. 
Note that we also perform identity filling to the "no gate" case.
Now with one SU(2) operation for every qubit during each cycle, we apply the U3 decomposition \cite{mckay2017efficient} to compile the SU(2) operation into two physical $\pi/2$ pulses and additional virtual Z gates, resulting in restless single-qubit gate cycles.
Since the virtual Z gates only affect the pulse phases, the entire waveform shape is almost identical given an arbitrary original circuit.
 
A direct consequence of this compiling scheme is that only the $\pi/2$ pulses are required to calibrate. More importantly, simultaneous calibration of these $\pi/2$ pulses immediately accounts for microwave signal crosstalk because all pulses have fixed shapes and hence fixed crosstalk influence; this substantially reduces the efforts for performing crosstalk compensation.
In addition, the timing of both single- and two-qubit pulses, is fixed during calibration and subsequent experiments. In this way, pulse distortions -- if not completely removable -- can be more consistent in different circuits; this makes the gate performance more predictable for long circuits.
Figure \ref{Fig2_circuit_complilation}\textbf{d} shows the circuits used for calibrating the controlled phase and single-qubit phase correction. With the isomorphous waveform method, their compiled waveforms are almost identical to each other and to the VQE circuits.
In addition, the method is compatible with randomized compiling~\cite{hashim2020randomized}, another powerful error mitigation technique. 
The method may require more single-qubit gates to be added to the circuits. However, this is unlikely to degrade the whole circuit fidelity due to the fact that the SU(2) squeeze operation can reduce the circuit depth and the fact that the single-qubit gate fidelities are typically much better than the two-qubit ones.

\begin{figure*}
  \centering
  \includegraphics[width=0.95\textwidth]{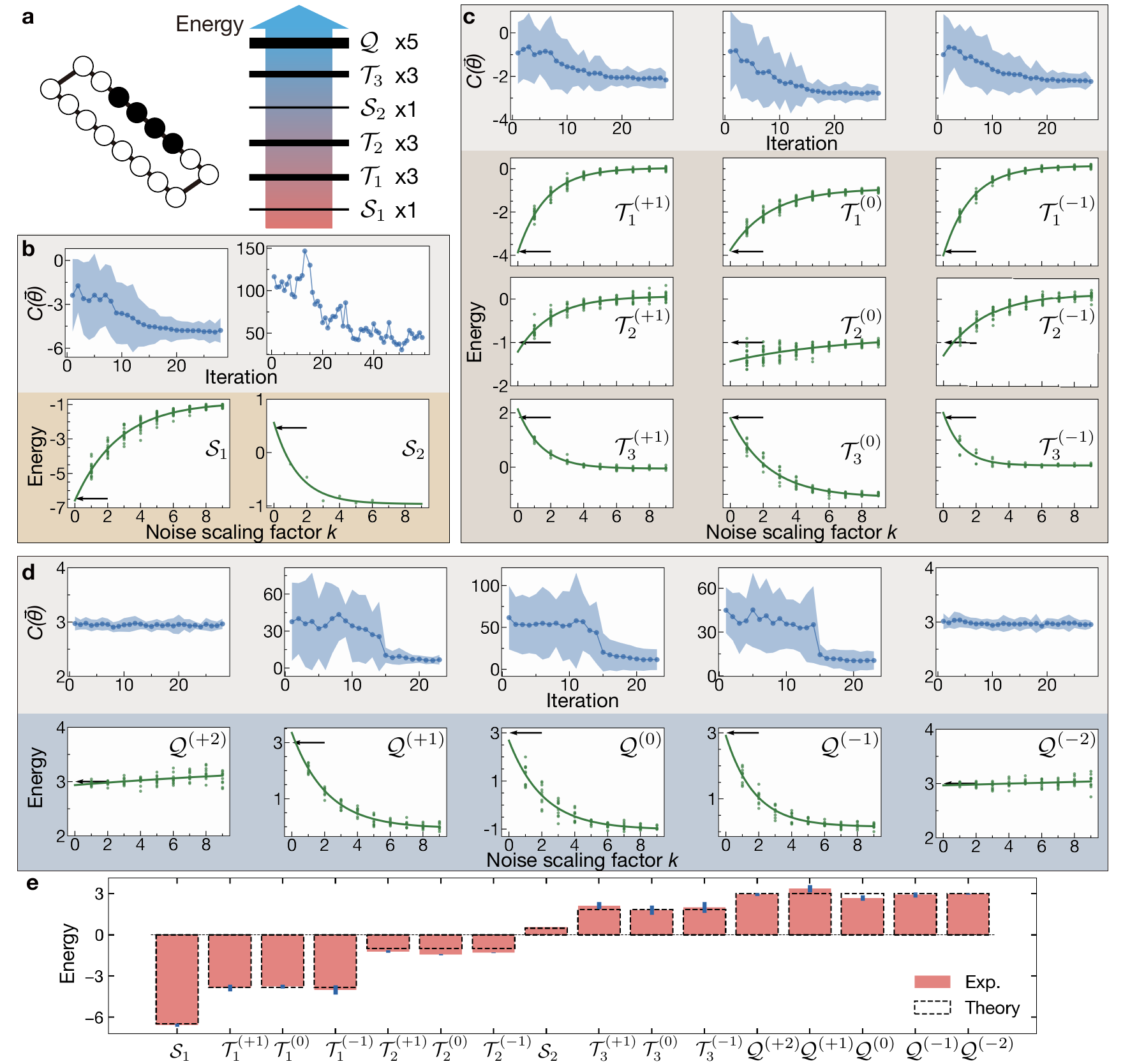}
  \caption[]
  {\textbf{Full spectroscopy of a 4-qubit Heisenberg spin chain.} 
  \textbf{a,} The left plot shows the device connectivity and the four active qubits (solid circles). The right plot sketches the spectrum of the 16 states of the 4-qubit Heisenberg spin chain which are grouped by degeneracy.
  \textbf{b-d,} VQE training and error mitigation for each of the 16 states including the singlets $\mathcal{S}_1$ and $\mathcal{S}_2$ (\textbf{b}), the triplets $\mathcal{T}_1$, $\mathcal{T}_2$ and $\mathcal{T}_3$ (\textbf{c}), and the quintuple $\mathcal{Q}$ (\textbf{d}). 
  The top panels (gray background) in each figure plot the training curves of the cost function $C(\vec{\theta})$. The blue dots are the mean values of the training curves with different initial parameters. The shades indicate twice the standard deviation.
  The bottom panels (colored background) plot the error mitigation results for different training outcomes (green dots) and the exponential fit to their mean (green line). The black arrows indicate the theoretical values. 
  Note that the triplet states with the same $s_z$ values share a common trained circuit and their energies can be obtained simply by switching the initialization circuit. 
  \textbf{e,} Comparison of the extrapolated zero-noise energies (red bars) and the theoretical values (dashed boxes) for all 16 eigenstates. The error bars (blue bars) denote plus/minus twice the standard deviation.}
  \label{Fig3_Heisenberg_N4}
\end{figure*}

After the processor is calibrated, we implement the full energy spectroscopy of a 4-qubit Heisenberg spin chain on the selected qubits (Fig.~\ref{Fig3_Heisenberg_N4}\textbf{a}).
Since all the eigenstates have definitive $s$, $s_z$, and $m$ numbers, the ground state $|E_1\rangle$ must be a global singlet (i.e.\ $s{=}s_z{=}0$) which we denote as $|\mathcal{S}_1 \rangle {=}|E_1\rangle$. Next in the spectrum are two triplets (i.e.\ $s{=}1$), each with degeneracy three corresponding to $s_z{=}0,\pm 1$, which we denote as $|\mathcal{T}_1^{(0,\pm 1)}\rangle=|E_{2,3,4}\rangle$ and $|\mathcal{T}_2^{(0,\pm 1)}\rangle=|E_{5,6,7}\rangle$ (the superscript denotes the $s_z$ number). The spectrum then follows with one more singlet $|\mathcal{S}_2 \rangle {=}|E_8\rangle$ and one triplet $|\mathcal{T}_3^{(0,\pm 1)}\rangle=|E_{9,10,11}\rangle$. The highest energy states are a quintuple (i.e.\ $s{=}2$) with degeneracy five corresponding to $s_z{=}0,\pm 1, \pm 2$, namely $| \mathcal{Q}^{(0,{\pm} 1, {\pm}2)} \rangle{=}|E_{12,\cdots,16}\rangle$.

To solve the singlet $| \mathcal{S}_1 \rangle$ which is the ground state, one can simply use the $s$-conserving circuit (3 parameters) in combination with mirror symmetry (to subtract 1 parameter) illustrated in Fig.~\ref{Fig1_schematic}\textbf{b}. Therefore, there are only 2 parameters per layer. We prepare the initial state $|\Phi\rangle{=}\ket{\psi^-} {\otimes} \ket{\psi^-}$, where $\ket{\psi^\pm}{=}(\ket{01} {\pm} \ket{10})/\sqrt{2}$. 
Since the initial state has a total spin $s{=}0$, the final state remains in the singlet manifold. A simple average energy minimization with single weight ($P{=}1$) in the cost function~\eqref{eq:SSVQE_cost} can estimate $E_{\mathcal{S}_1}$. The result converges to a precise value for a circuit with $L{=}2$ layers. Figure \ref{Fig3_Heisenberg_N4}\textbf{b} shows the convergence of the cost function versus optimization iterations and the energy estimation as a function of noise scaling factor $k$. The ground state energy extrapolated at $k{=}0$ yields $E_{\mathcal{S}_1}{=} (-6.570 \pm 0.059)$ (theory: $-6.464$) assuming $J{=}1$. 
Note that our approach requires only 4 parameters to train, while the widely used hardware efficient ansatz~\cite{kandala2017hardware} requires a total of 3 layers and 32 parameters. As shown in Ref.~\cite{Lyu2023symmetryenhanced}, the situation gets worse as the system size or the number of target states in SSVQE increases.

For singlet $| \mathcal{S}_2 \rangle$, we resort to the SSVQE method by training $| \mathcal{S}_1 \rangle$ and $| \mathcal{S}_2 \rangle$ simultaneously with double initial states ($P{=}2$ in the cost function). Unlike the ground state, producing two orthogonal initial states with $s{=}0$ requires long-distance controlled-NOT gates which are not available in our device. We instead use the $s_z$-conserving circuit with $L{=}3$ layers and two initial states $\ket{\Phi_1}{=}\ket{0101}$ and $\ket{\Phi_2}{=}\ket{1010}$. To ensure that the final states fall into the $s{=}0$ manifold, an additional penalty term $\langle S_\mathrm{tot}^2 \rangle$ is included in the cost function (see Supplementary Information for details). Due to the complex nature of the cost function, the training becomes much harder than the ground state and takes about 24 hours until convergence. We only obtain one successful convergence (Fig.~\ref{Fig3_Heisenberg_N4}\textbf{b}) out of a few trials as the processor performance fluctuates over time. 
Interestingly, we find that in this case, the hybrid optimization of Nelder-Mead (first $14$ iterations) followed by Adam performs better than either of them separately.

For the three triplet groups $\mathcal{T}_{1,2,3}$, the eigenstates can be efficiently solved with fewer training tasks by identifying symmetries in these states and choosing proper initial states. 
To understand this, let us focus on the case of $s_z{=}0$. We use the total spin $s$-conserving circuit with mirror symmetry with $L{=}2$ layers. For the initial state $\ket{\Phi_1}{=}(\ket{0101}{-}\ket{1010})/\sqrt{2}$ which is determined by symmetry numbers $s{=}1,s_z{=}0,m{=}-1$, minimizing the average energy yields the lowest eigenstate in the same-symmetry manifold, which is exactly $| \mathcal{T}_{1}^{(0)} \rangle$. 
Then applying the same trained circuit on $\ket{\Phi_2}{=}(\ket{0011}{-}\ket{1100})/\sqrt{2}$ which is orthogonal to $\ket{\Phi_1}$ but shares the same symmetries immediately generates $| \mathcal{T}_{3}^{(0)} \rangle$. 
Similarly, applying the same trained circuit on $\ket{\Phi_3}{=}(\ket{0110}{-}\ket{1001})/\sqrt{2}$ which is orthogonal to $\ket{\Phi_1}$ and $\ket{\Phi_2}$ but with $m{=}{+}1$ unambiguously results in $| \mathcal{T}_{2}^{(0)} \rangle$.
In this manner, all three triplet states $s_z{=}0$ share a common trained circuit, and error mitigation is conducted separately (Fig.~\ref{Fig3_Heisenberg_N4}\textbf{c}).
The same treatment also works for the manifold of $s_z{=}{\pm} 1$.
In Supplementary Information, we give the exact form of all the initial states and their corresponding circuits. 

The last manifold of the spectrum is a quintuple with $s{=}2$. Two eigenstates with $s_z{=}{\pm}2$ are easy to obtain as all the spins are aligned. Therefore, we focus on generating the three eigenstates with $s_z{=}0,{\pm}1$. 
Generating initial states with total spin $s{=}2$ and $s_z{=}0,{\pm}1$ is not easy without long-distance controlled-NOT gates. 
Instead, we add the penalty term $\langle (S_\mathrm{tot}^2 - 6)^2 \rangle$ to the cost function with $P{=}1$ to guarantee the convergence to the $s{=}2$ manifold. We use the $s_z$-conserving circuit with mirror symmetry and the initial states $\ket{0101}$ ($s_z{=}0$), $\ket{0111}$ ($s_z{=}{-}1$), $\ket{0001}$ ($s_z{=}1$). The results are shown in Fig.~\ref{Fig3_Heisenberg_N4}\textbf{d}. 
Again, we find that similar to the SSVQE simulation of $| \mathcal{S}_2 \rangle$ (Fig.~\ref{Fig3_Heisenberg_N4}\textbf{b}), the hybrid optimization of the gradient-free Nelder-Mead (first $14$ iterations) and gradient-based Adam method gives the best convergence. In fact, when the cost function contains several terms, the potential landscape becomes complex with multiple local minima. It is likely that a few iterations by a gradient-free optimizer, such as Nelder-Mead, bring the cost function to the vicinity of its global minimum whereas a following gradient-based optimizer, such as Adam, facilitates the convergence. This can be seen from the immediate drop in the cost function when switching from Nelder-Mead to Adam at the 15th iteration in all the relevant cases. 
The estimated energy eigenvalues of all the $16$ states are compared with the theory in Fig.~\ref{Fig3_Heisenberg_N4}\textbf{e}. The average deviation is about 0.13, validating our approach for obtaining the full variational spectroscopy.

\begin{figure*}
  \centering
  \includegraphics[width=0.95\textwidth]{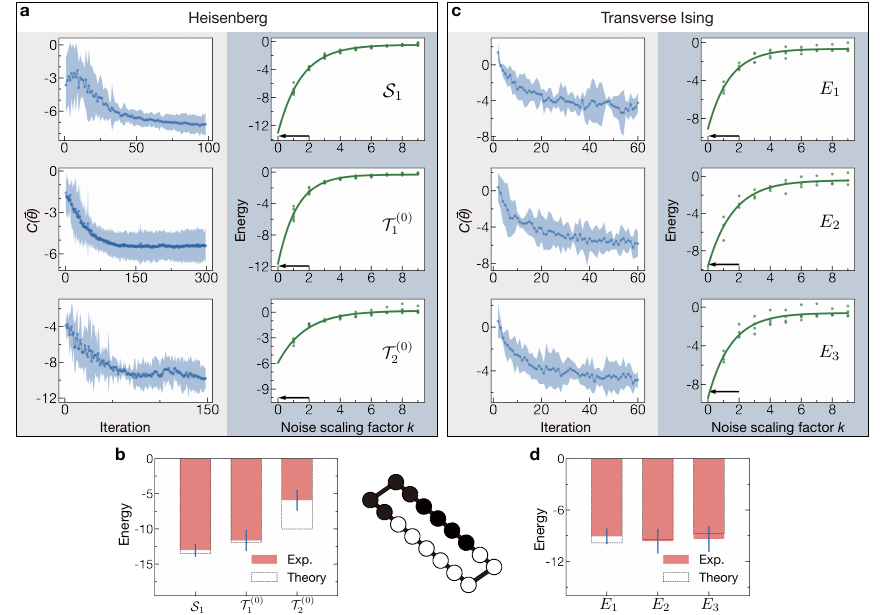}
  \caption[]{
  \textbf{Low-energy spectroscopy of an 8-qubit Heisenberg and transverse Ising chain.}  
  \textbf{a,} VQE training (left panels) and error mitigation (right panels) for the three lowest eigenstates for the 8-qubit chain with the Heisenberg interactions.
  \textbf{b,} Comparison of the extrapolated zero-noise energies (red bars) and the theoretical values (dashed box) of the Heisenberg model. The error bars (blue bars) are plus/minus twice the standard deviation from the measured values.
  \textbf{c,} Same as (\textbf{a}) but with the transverse Ising interactions.
  \textbf{d,} Same as (\textbf{b}) but for the Ising model.
  The center panel at bottom shows the eight active qubits (solid circles).}
  \label{Fig4_Heisenberg_Ising_N8}
\end{figure*}

To show the scalability of our approach, we perform low-energy spectroscopy for an 8-qubit Heisenberg Hamiltonian. We consider estimating the three lowest eigenenergies, namely $E_{\mathcal{S}_{1}}$, $E_{\mathcal{T}_{1}}$ and $E_{\mathcal{T}_{2}}$. For the ground state $\mathcal{S}_{1}$, we use the $s$-conserving circuit with $L{=}3$ layers, and the initial state $|\Phi\rangle{=}\ket{\psi^-}^{\otimes 4}$. The training of the cost function ($P{=}1$) and the error mitigation results are shown in Fig.~\ref{Fig4_Heisenberg_Ising_N8}\textbf{a}. 
For the first excited state $\mathcal{T}_{1}^{(0)}$, we use the $s$-conserving circuit with $L{=}4$ layers and the initial state $|\Phi\rangle{=}\ket{\psi^-}\ket{\psi^+}\ket{\psi^-}\ket{\psi^-}$. Remarkably, there are 84 CNOT gates in one cycle of the parameterized circuit and a total of 792 CNOT gates in the error mitigation circuit with scaling factor $k{=}9$. The consistency in the measured data (middle panels of Fig.~\ref{Fig4_Heisenberg_Ising_N8}\textbf{a}) manifests the consistency in our gate performance in large-scale circuits, owing to the isomorphous waveform technique.
For other choices of $s_z$, one can simply change $\ket{\psi^+}$ to $\ket{00}$ or $\ket{11}$.  Again, if one is interested to get both 
$\mathcal{T}_{1}^{(0)}$ and $\mathcal{T}_{2}^{(0)}$, simultaneously, one has to use SSVQE with $P{=}2$. While one of the initial states is the same as before, the other one can be $|\Phi\rangle=\ket{\psi^-}\ket{\psi^-}\ket{\psi^+}\ket{\psi^-}$. The results are shown in the lower panels of Fig.~\ref{Fig4_Heisenberg_Ising_N8}\textbf{a}, where the $\mathcal{T}_{1}^{(0)}$ trace is dropped for clarity.  
The three estimated eigenenergies and their theoretical values are compared in Fig.~\ref{Fig4_Heisenberg_Ising_N8}\textbf{b}. While $E_{\mathcal{S}_{1}}$ and $E_{\mathcal{T}_{1}}$ can be estimated very precisely, the energy $E_{\mathcal{T}_{2}}$ shows a larger error. This is because a circuit with $L{=}4$ layers is not capable to target two eigenstates simultaneously with high precision.

The key advantage of programmable digital quantum simulators is their capability in simulating different Hamiltonian models.
To show the versatility, we also perform on the same device the variational spectroscopy for the transverse-field Ising Hamiltonian  
\begin{equation}
  H = J \sum_{i=1}^{N-1} \sigma_x^i \sigma_x^{i+1} + h \sum_i \sigma_z^i \;,
  \label{eq:transverse_Ising}
\end{equation}
where $J$ is the coupling strength and $h$ is the transverse field strength. This system has a critical point at $h{=}J$ where the ground state becomes highly entangled, and thus its variational simulation is expected to be harder~\cite{lyu2022variational,koffel2012Entanglement}. The Hamiltonian commutes with two operators, namely the mirror operator $\mathcal{M}$ and the parity operator $\Pi_{z}{=}\bigotimes_{i=1}^N \sigma_z^i$, which imply that every eigenstate has two symmetry numbers, $m{=}{\pm}1$ (i.e.\ $\mathcal{M}\ket{E_j}{=}m\ket{E_j}$) and $\pi_z{=}{\pm}1$ (i.e.\ $\Pi_{z}\ket{E_j}{=}\pi_z\ket{E_j}$). Hence, We can denote the state as $|E_j^{(m, \pi_z)}\rangle$.
For simulating the transverse Ising Hamiltonian, the two-qubit operation $\mathcal{N}(\theta)$ is replaced with 
\begin{equation}
  \mathcal{N}_{\mathrm{I}}(\theta) = e^{-i \theta \sigma_x^i \sigma_x^{i+1}/2}  \;.
\end{equation}
The circuit used to construct $\mathcal{N}_{\mathrm{I}}(\theta)$ is depicted in Fig.~\ref{Fig1_schematic}\textbf{b}. To ensure the preservation of the $\Pi_{z}$ symmetry, the single-qubit operations should all be $z$-rotations. Furthermore, mirror symmetry is readily preserved by choosing parameters symmetrically with respect to the middle of the chain. The three lowest energy eigenstates are denoted as $|E_1^{(1, 1)}\rangle$, $|E_2^{(-1, -1)}\rangle$ and $|E_3^{(1, -1)}\rangle$. These three eigenstates possess symmetry numbers that allow their independent generation through the appropriate choice of the initial state and a cost function with only $P{=}1$. 
We simulate these three eigenstates for both 4-qubit and 8-qubit Ising chains, presenting the result for the 8-qubit case here. The 4-qubit results and other experimental details can be found in Supplementary Information. 
We employ the $\pi_{z}$-conserving circuit ($L{=}3$) with mirror symmetry and the initial state $|\Phi\rangle{=}\ket{000} |\phi_{(4,5)}\rangle \ket{000}$, where $|\phi_{(4,5)} \rangle$ represents the quantum state of qubits 4 and 5. In order to satisfy the desired symmetries, $|\phi_{(4,5)} \rangle$ is chosen to be one of the Bell states $\ket{\phi^+}{=}(\ket{00}{+}\ket{11})/\sqrt{2}$, $\ket{\psi^-}$ and $\ket{\psi^+}$ for estimating $E_1$, $E_2$ and $E_3$, respectively. The training and error mitigation results are shown in Fig.~\ref{Fig4_Heisenberg_Ising_N8}\textbf{c}. A good agreement between the measured energies and the theoretical values is observed (Fig.~\ref{Fig4_Heisenberg_Ising_N8}\textbf{d}).\\  


In our study, we present an experimental demonstration of multi-level variational spectroscopy using a programmable superconducting quantum simulator. Our findings highlight the resource efficiency afforded by symmetry and the ability to target a greater number of states through the implementation of the subspace search method and the inclusion of penalty terms. However, a significant challenge is the circuit depth overhead, which is inherent to the SSVQE algorithm when the number of target states increases. Furthermore, the limited qubit connectivity of our one-dimensional simulator imposes constraints on generating appropriate initial states. Both increasing qubit connectivity by utilizing simulators based on two-dimensional qubit arrays or implementing the protocol on alternative platforms such as ion traps should address these limitations. 
Additionally, we discover that a hybrid optimization strategy, which combines gradient-free Nelder-Mead with gradient-based Adam, offers faster convergence when penalty terms are present in the cost function. This suggests that in a disrupted potential landscape caused by relatively strong penalty terms, Nelder-Mead is less sensitive to local minima, guiding the cost function toward its global minimum. At this point, the Adam optimizer can rapidly converge.


\bibliographystyle{naturemag}
\bibliography{References}

\newpage

\noindent\textbf{Methods}

~\\

\noindent\textbf{VQE training.}
To ensure that our convergence is accurate, we use a restricted range of the initial parameters, which are uniformly sampled from -$\pi{/}2$ to $\pi{/}2$. These parameters are used as a starting point for the VQE training procedure. The variational circuit with initial parameters is then compiled into isomorphic forms as described in the main text and further discussed in Supplementary Information. At each iteration, the cost function is measured by measuring relevant correlation functions. To ensure accuracy,  each measurement is repeated around $6000$ times. Then by feeding the cost function into our classical optimizer, we obtain a new set of parameters $\vec{\theta}$.  We consider three different ways for classical optimization, namely gradient-free Nelder-Mead, gradient-based Adam, and a hybrid of both. In the Nelder-Mead algorithm, the initial simplex is uniformly sampled from -$\pi{/}2$ to $\pi{/}2$. In the Adam optimizer, the initial parameters are also uniformly sampled from -$\pi{/}2$ to $\pi{/}2$.  The configuration of Adam optimizer is chosen to be $\alpha{=}0.1$ for the learning rate, $\beta_1{=}0.9$ for the exponential decay rate of the 1st-moment, $\beta_2{=}0.999$ for the exponential decay rate of the 2nd-moment, and finally $\epsilon{=}10^{-8}$  for numerical stability. In the hybrid optimization, we utilize the Nelder-Mead for ${\sim}15$ iterations to obtain a rough training result, and then use Adam until the training converges. 

In most cases, the Nelder-Mead algorithm performed better than the other two, See Supplementary Information for specifying the best optimizer for every single case. However, when the cost function contains several terms, e.g. in the SSVQE with $P{>}1$ or in the presence of penalizing terms,  the best optimization method is often the hybrid method. This might be due to the presence of multiple local minima in such a complex potential landscape. In such cases, Nelder-Mead brings the cost function to the vicinity of its global minimum but it is slow to make the final convergence  towards the exact location of the global minimum. However, the gradient-based Adam optimizer operates well when the cost function is close to its global minimum. As a result, for complex multi-term cost functions, hybrid optimizations seem to provide faster convergence.



~\\
\noindent\textbf{Error mitigation.}
In our experiment, we utilize the initialization circuits to prepare initial states that satisfy the corresponding $S_z$ value and ansatz circuits to conserve $S_z$. To ensure that the conservation of $S_z$ is maintained, we remove the data that fails to meet this criterion.

To mitigate errors, we employ the zero-noise extrapolation (ZNE) technique, which involves proportionally increasing the entire circuit error.
For the initialization circuit $U_{\mathrm{I}}$, we increase the circuit error by a factor of $k$ (where $k{=}2n{+}1$ is an odd number) by replacing the circuit $U_{\mathrm{I}}$ with $(U_{\mathrm{I}} U_{\mathrm{I}}^\dagger)^{n} U_{\mathrm{I}}$. However, if $k{=}2n$, we take the average of $(U_{\mathrm{I}} U_{\mathrm{I}}^\dagger)^{n-1} U_{\mathrm{I}}$ and $(U_{\mathrm{I}} U_{\mathrm{I}}^\dagger)^{n} U_{\mathrm{I}}$ to obtain the desired results.
For the ansatz circuits, we can increase the circuit error by a factor of $k$ by substituting the original $N(\theta)$ with $N(\theta/k)^k$.
Such a protocol allows us to utilize even noise scaling factors, improving the statistics during extrapolation given a limited circuit length.

~\\
\noindent\textbf{ Device and setup.}
The sample (10mm-by-10mm in size) is made of aluminum on a sapphire substrate and packaged in an aluminum sample enclosure. The sample is mounted inside a Bluefors LD400 dilution refrigerator at a base temperature of 12~mK and shielded with two layers of cryo-perm cans.

The qubits have alternating frequencies between 3.8~GHz and 4.2~GHz, designed for suppressing residual ZZ interactions. The couplers idle at their maximum frequencies of 6.4~GHz and their couplings to the qubits are about 100~MHz. The readout resonators designed at $\sim$7~GHz have photon loss rates of 1~MHz and a dispersive shift of 0.6~MHz. Eight resonators share a common transmission line to which a 20~fF capacitor is added at the input end to prevent photons from entering this port.

The XY microwave signal for single-qubit gates is synthesized by up-converting intermediate-frequency signals with a single-tone microwave carrier using an IQ mixer. To ensure high signal quality without reflections and spurious frequencies, an isolator and a band-pass filter are used. The fast Z signal for two-qubit gates is combined with the XY signal using a diplexer at room temperature. The combined signal is then attenuated and filtered at multiple cold stages of the DR for noise suppression. A custom-made low-pass IR filter is added to each control line at the mixing-chamber stage which attenuates signals at the qubit frequency by approximately 25 dB while allowing low-frequency signals to pass through. 

The multiplexed readout signal is generated from up-conversion using an IQ mixer, filtered by a 7-GHz band-pass filter, and attenuated at different temperature stages before entering the device. The output signal is sequentially amplified by a HEMT amplifier at 4K and an amplifier at room temperature before being down-converted and demodulated. To block noise from higher temperature stages, circulators and filters are used at the output port. More device and setup information can be found in Supplementary Information.

\section{}

\noindent\textbf{Data availability}\\
Data that support the plots within this paper and other findings of this study are available from the corresponding author upon reasonable request.\\

\noindent\textbf{Acknowledgements}\\
This work was supported by the Key-Area Research and Development Program of Guangdong Province (2018B030326001), the National Natural Science Foundation of China (U1801661), the Guangdong Innovative and Entrepreneurial Research Team Program (2016ZT06D348), the Guangdong Provincial Key Laboratory (2019B121203002), the Science, Technology, and Innovation Commission of Shenzhen Municipality (KYTDPT20181011104202253), the Shenzhen-Hong Kong Cooperation Zone for Technology and Innovation (HZQB-KCZYB-2020050) and the NSF of Beijing (Z190012).
A.B. acknowledges support from the National Key R\&D Program of China (2018YFA0306703), the National Science Foundation of China (12050410253, 92065115, 12274059), the Innovation Program for Quantum Science and Technology (ZD0301703), and the Ministry of Science and Technology of China (QNJ2021167001L).\\

\noindent\textbf{Author contributions}\\
Z.H., C.L., A.B., and F.Y. designed the experiment, analyzed the data, and wrote the manuscript. 
Z.H. conducted the measurements. 
Z.H., Y.Z., J.Y., J.C., L.H., and F.Y. designed the device. 
Y.Z., H.J., L.N., W.W., and L.Z. performed sample fabrication. 
W.N., Z.Y., and Z.Z. supported with the software. 
C.H., L.H., J.L., and D.T. assisted with the measurement setup.  
A.B., S.L., F.Y., and D.Y. supervised the project. 
All authors discussed the results and contributed to revising the manuscript and the Supplementary Information. 
All authors contributed to the experimental and theoretical infrastructure to enable the experiment.



\end{document}


\title{Supplementary Information for “Multi-Level Variational Spectroscopy using a Programmable Quantum Simulator” }

\newcommand{\SIQSE}{\affiliation{1}{Shenzhen Institute for Quantum Science and Engineering, Southern University of Science and Technology, Shenzhen, Guangdong, China}}
\newcommand{\IQA}{\affiliation{2}{International Quantum Academy, Shenzhen, Guangdong, China}}
\newcommand{\GDKL}{\affiliation{3}{Guangdong Provincial Key Laboratory of Quantum Science and Engineering, Southern University of Science and Technology, Shenzhen, Guangdong, China}}
\newcommand{\DPHY}{\affiliation{4}{Department of Physics, Southern University of Science and Technology, Shenzhen, Guangdong, China}}
\newcommand{\HFNL}{\affiliation{5}{Shenzhen Branch, Hefei National Laboratory, Shenzhen, China}}
\newcommand{\UESTC}{\affiliation{6}{Institute of Fundamental and Frontier Sciences,
University of Electronic Science and Technology of China, Chengdu, China}}

\author{Zhikun Han}
\thanks{These authors have contributed equally to this work.}
\affiliation{\SIQSE}\affiliation{\IQA}\affiliation{\GDKL}

\author{Chufan Lyu}
\thanks{These authors have contributed equally to this work.}
\affiliation{\UESTC}

\author{Yuxuan Zhou}
\thanks{These authors have contributed equally to this work.}
\affiliation{\SIQSE}\affiliation{\IQA}\affiliation{\GDKL}\affiliation{\DPHY}

\author{Jiahao Yuan}
\affiliation{\SIQSE}\affiliation{\IQA}\affiliation{\GDKL}\affiliation{\DPHY}
\author{Ji Chu}
\affiliation{\SIQSE}\affiliation{\IQA}\affiliation{\GDKL}
\author{Wuerkaixi Nuerbolati}
\affiliation{\SIQSE}\affiliation{\IQA}\affiliation{\GDKL}
\author{Hao Jia}
\affiliation{\SIQSE}\affiliation{\IQA}\affiliation{\GDKL}
\author{Lifu Nie}
\affiliation{\SIQSE}\affiliation{\IQA}\affiliation{\GDKL}
\author{Weiwei Wei}
\affiliation{\SIQSE}\affiliation{\IQA}\affiliation{\GDKL}
\author{Zusheng Yang}
\affiliation{\SIQSE}\affiliation{\IQA}\affiliation{\GDKL}
\author{Libo Zhang}
\affiliation{\SIQSE}\affiliation{\IQA}\affiliation{\GDKL}
\author{Ziyan Zhang}
\affiliation{\SIQSE}\affiliation{\IQA}\affiliation{\GDKL}

\author{Chang-Kang Hu}
\affiliation{\SIQSE}\affiliation{\IQA}\affiliation{\GDKL}
\author{Ling Hu}
\affiliation{\SIQSE}\affiliation{\IQA}\affiliation{\GDKL}
\author{Jian Li}
\affiliation{\SIQSE}\affiliation{\IQA}\affiliation{\GDKL}
\author{Dian Tan}
\affiliation{\SIQSE}\affiliation{\IQA}\affiliation{\GDKL}

\author{Abolfazl Bayat}
\email{abolfazl.bayat@uestc.edu.cn}
\affiliation{\UESTC}

\author{Song Liu}
\email{lius3@sustech.edu.cn}
\affiliation{\SIQSE}\affiliation{\IQA}\affiliation{\GDKL}\

\author{Fei Yan}
\email{yanfei@baqis.ac.cn}
\altaffiliation[Present address: ]{Beijing Academy of Quantum Information Sciences, Beijing, China}
\affiliation{\SIQSE}\affiliation{\IQA}\affiliation{\GDKL}

\author{Dapeng Yu}
\email{yudp@sustech.edu.cn}
\affiliation{\SIQSE}\affiliation{\IQA}\affiliation{\GDKL}\affiliation{\DPHY}

\maketitle


\section*{Initial state construction}

In order to approximate the desired eigenstate with the symmetry-preserving ansatzes in VQE, the initialization needs to be carefully chosen for the corresponding symmetry. For the Heisenberg Hamiltonian with system size $N{=}4$, the two singlet eigenvectors $| \mathcal{S}_{1,2} \rangle$ share the symmetry of total spin $s{=}0$, $z$-magnetization $s_z{=}0$, and mirror symmetry $m{=}1$. In order to target $| \mathcal{S}_{1,2} \rangle$, two initial states with these symmetry numbers are required. These states can be obtained by Clebsch-Gordan decomposition as $| \psi^- \rangle \otimes | \psi^- \rangle$ and $\frac{\sqrt{3}}{6}(\ket{0101} + \ket{1010} + \ket{0110} + \ket{1001} - 2\ket{1100} - 2\ket{0011})$. 
The initialization circuit for the former can be easily obtained which is depicted in Table \ref{tab:QinfoN4heis1}. However, the latter requires a non-trivial initialization circuit with long-distance controlled-not gates which is not available in our device. Therefore, we use initial states $\ket{0101}$ and $\ket{1010}$, see their corresponding quantum circuits in Table \ref{tab:QinfoN4heis1}, with $S_z$-conserving circuit to approximate $\mathcal{S}_{2}$ using SSVQE. 

The three triplets $| \mathcal{T}_{1,2,3}^{(0,{\pm}1)} \rangle$ share the symmetry of total spin $s{=}1$. Each of these eigenstates is triply degenerate with $s_z{=}0,{\pm}1$. The corresponding mirror symmetry are $m{=}{-}1$ for $| \mathcal{T}_{1,3}^{(0,{\pm}1)} \rangle$, and $m{=}{+}1$ for $| \mathcal{T}_{2}^{(0,{\pm}1)} \rangle$. Therefore, for each given $s_z$, we use the following initial states:
$$
\mathrm{For}~{s_z{=}0:}\left\{
\begin{aligned}
    & \ket{\Phi_1} =& (\ket{0101}-\ket{1010})/\sqrt{2}\\
    &\ket{\Phi_2} =& (\ket{0110}-\ket{1001})/\sqrt{2} \\
    &\ket{\Phi_3} =& (\ket{0011}-\ket{1100})/\sqrt{2}   
\end{aligned}
\right.$$
$$\mathrm{For}~{s_z{=}{+}1:}\left\{
\begin{aligned}
    & \ket{\Phi_1} =& (\ket{0001}-\ket{0010}+\ket{0100}-\ket{1000})/2\\
    &\ket{\Phi_2} =& (\ket{0001}-\ket{0010}-\ket{0100}+\ket{1000})/2 \\
    &\ket{\Phi_3} =& (\ket{0001}+\ket{0010}-\ket{0100}-\ket{1000})/2
\end{aligned}
\right. $$
$$\mathrm{For}~{s_z{=}{-}1:}\left\{
\begin{aligned}
    & \ket{\Phi_1} =& (\ket{0111}-\ket{1011}+\ket{1101}-\ket{1110})/2\\
    &\ket{\Phi_2} =& (\ket{0111}-\ket{1011}-\ket{1101}+\ket{1110})/2 \\
    &\ket{\Phi_3} =& (\ket{0111}+\ket{1011}-\ket{1101}-\ket{1110})/2   
\end{aligned}
\right.
$$
The quantum circuits for generating these states are shown in Table \ref{tab:QinfoN4heis1} and Table \ref{tab:QinfoN4heis2}.

For the quintuple $| \mathcal{Q}^{(0,{\pm}1,{\pm}2)} \rangle$, which is determined by total spin $s{=}2$, there are five degenerate eigenstates given by $s_z{=}{-}2,{-}1,0,{+}1,{+}2$. To generate these eigenstates, we use a $s_z$-conserving circuit together with a penalty term added to the cost function, see the main text for more details. Therefore, the initial states for these five eigenstates are simply set to be $\ket{1111},\ket{0111},\ket{0101}, \ket{0001}, \ket{0000}$. The corresponding initialization circuits are shown in Table \ref{tab:QinfoN4heis2}.

In the case of Heisenberg Hamiltonian with system size $N{=}8$, we simulate the $3$ lowest eigenenergies. Similar to the $N{=}4$ case, the ground state $| \mathcal{S}_1 \rangle$ is a global singlet with total spin $s{=}0$, $z$-magnetization $s_z{=}0$, and mirror symmetry $m{=}1$. The initial state for the ground state is $\ket{\psi^-}^{\otimes 4}$. For the first excited state $|\mathcal{T}^{(0)}_1 \rangle$ with $s{=}{1},s_z{=}{0},m{=}-1$, the initial state is set to be $\ket{\psi^-}\ket{\psi^+}\ket{\psi^-}\ket{\psi^-}$. Since the second excited state $|\mathcal{T}^{(0)}_2 \rangle$ share the same symmetry numbers as the first excited state, its initial state can similarly choose as $\ket{\psi^-}\ket{\psi^-}\ket{\psi^+}\ket{\psi^-}$. The initialization circuits for these three initial states are shown in Table \ref{tab:QinfoN8heis}. For other eigenstates with $s_z{=}{\pm}1$, one can simply replace $\ket{\psi^+}$ with $\ket{00}$ or $\ket{11}$. The circuits are not shown, as they can be easily obtained by replacing the circuit for $\ket{\psi^+}$ with the circuit for $\ket{00}$ or $\ket{11}$.

The symmetry numbers for the low-energy spectrum of Ising Hamiltonian with 4-qubit and 8-qubit systems are the same. Namely, the ground state $\ket{E_1}$ is determined by $z{=}1$ for the parity operator $\Pi_z$, and $m{=}{+}1$ for mirror symmetry $\mathcal{M}$. Hence, its initial state for both system sizes $N{=}4$ and $N{=}8$ is set to be $\ket{0}^{\otimes \frac{N{-}2}{2}} \ket{\phi^+} \ket{0}^{\otimes \frac{N{-}2}{2}}$. The second eigenstate, $\ket{E_2}$, exhibits $z{=}{-}1$ and $m{=}{-}1$ symmetries, with the initial state given by $\ket{0}^{\otimes \frac{N-2}{2}} \ket{\psi^-} \ket{0}^{\otimes \frac{N-2}{2}}$. The third eigenstate, $\ket{E_3}$, has $z{=}{-}1$ and $m{=}{+}1$ symmetries, and its initial state is expressed as $\ket{0}^{\otimes \frac{N-2}{2}} \ket{\psi^+} \ket{0}^{\otimes \frac{N-2}{2}}$. The initialization circuits for these three eigenstates are shown in Table \ref{tab:QinfoN4Ising} and Table \ref{tab:QinfoN8Ising}.

\begin{table*}[htbp]
\centering
\caption{Circuits for the 4-qubit Heisenberg model}
\begin{ruledtabular}
\begin{tabular}{c|ccccccccc}

\textbf{State}& \textbf{Initialization Circuit} & \textbf{Ansatz layer} & \textbf{\# of layers} \\ \hline

$\mathcal{S}_1$ & 
\begin{tikzcd}[row sep=.1cm, column sep=.1cm]
    \lvert 0\rangle& \gate[1]{X_\pi} & \gate{H} & \ctrl{1}  & \qw\\
    \lvert 0\rangle& \gate{X_\pi} & \qw &  \targ{} & \qw\\
    \lvert 0\rangle& \gate{X_\pi} & \gate{H} & \ctrl{1}  & \qw \\
    \lvert 0\rangle& \gate{X_\pi} & \qw & \targ{}  & \qw 
\end{tikzcd} 
&
 \begin{tikzcd}[row sep=.1cm, column sep=.1cm]
& \gate[2]{\mathcal{N}_{\mathrm{H}}(\theta_{1})} & \qw &  \qw &\\
& \qw & \gate[2]{\mathcal{N}_{\mathrm{H}}(\theta_{2})} &  \qw &\\
& \gate[2]{\mathcal{N}_{\mathrm{H}}(\theta_{1})} & \qw &  \qw &\\
& \qw & \qw  & \qw &
\end{tikzcd}
& 2 \\ \hline

$\mathcal{S}_2$ & 
\begin{tikzcd}[row sep=.1cm, column sep=.1cm]
    \lvert 0\rangle& \gate[1]{X_\pi} & \qw\\
    \lvert 0\rangle& \qw & \qw &  \qw \\
    \lvert 0\rangle& \gate{X_\pi} &  \qw   \\
    \lvert 0\rangle& \qw & \qw 
\end{tikzcd},
\begin{tikzcd}[row sep=.1cm, column sep=.1cm]
    \lvert 0\rangle& \qw & \qw\\
    \lvert 0\rangle& \gate{X_\pi} & \qw &  \qw \\
    \lvert 0\rangle& \qw &  \qw   \\
    \lvert 0\rangle& \gate{X_\pi} & \qw 
\end{tikzcd} 
&
 \begin{tikzcd}[row sep=.1cm, column sep=.1cm]
& \gate[2]{\mathcal{N}_{\mathrm{H}}(\theta_{1})} & \qw & \gate{Z(\theta_{3})} & \qw \\
& \qw & \gate[2]{\mathcal{N}_{\mathrm{H}}(\theta_{2})} & \gate{Z(\theta_{4})} & \qw \\
& \gate[2]{\mathcal{N}_{\mathrm{H}}(\theta_{1})} & \qw & \gate{Z(\theta_{4})} & \qw \\
& \qw & \qw & \gate{Z(\theta_{3})} & \qw
\end{tikzcd}
& 3 \\ \hline

$\mathcal{T}_1^{(+1)}$   &
\begin{tikzcd}[row sep=.1cm, column sep=.1cm]
    \lvert 0\rangle& \gate[1]{X_\pi
    } & \qw & \qw & \targ{}  & \qw & \qw & \qw & \qw & \qw & \qw & \qw & \qw \\
    \lvert 0\rangle& \gate{Y_a} & \gate{X_\pi} & \gate{Y_{-a}}& \ctrl{-1} & \ctrl{1} & \qw & \targ{} & \qw & \qw & \qw & \qw  & \gate{Z_\pi} \\
    \lvert 0\rangle& \qw & \qw & \qw & \gate{Y_b} & \targ{} & \gate{Y_{-b}} & \ctrl{-1} & \ctrl{1} & \qw & \targ{}  & \qw &\qw \\
    \lvert 0\rangle& \qw & \qw & \qw & \qw & \qw & \qw & \gate{Y_c} & \targ{} & \gate{Y_{-c}} & \ctrl{-1} & \qw & \gate{Z_\pi}
\end{tikzcd}
&
 \begin{tikzcd}[row sep=.1cm, column sep=.1cm]
& \gate[2]{\mathcal{N}_{\mathrm{H}}(\theta_{1})} & \qw &  \qw &\\
& \qw & \gate[2]{\mathcal{N}_{\mathrm{H}}(\theta_{2})} &  \qw &\\
& \gate[2]{\mathcal{N}_{\mathrm{H}}(\theta_{1})} & \qw &  \qw &\\
& \qw & \qw  & \qw &
\end{tikzcd}
& 2 \\ \hline

$\mathcal{T}_2^{(+1)}$  &
\begin{tikzcd}[row sep=.1cm, column sep=.1cm]
    \lvert 0\rangle& \gate[1]{X_\pi
    } & \qw & \qw & \targ{}  & \qw & \qw & \qw & \qw & \qw & \qw & \qw & \qw \\
    \lvert 0\rangle& \gate{Y_a} & \gate{X_\pi} & \gate{Y_{-a}}& \ctrl{-1} & \ctrl{1} & \qw & \targ{} & \qw & \qw & \qw & \qw  & \gate{Z_\pi} \\
    \lvert 0\rangle& \qw & \qw & \qw & \gate{Y_b} & \targ{} & \gate{Y_{-b}} & \ctrl{-1} & \ctrl{1} & \qw & \targ{}  & \qw &\gate{Z_\pi} \\
    \lvert 0\rangle& \qw & \qw & \qw & \qw & \qw & \qw & \gate{Y_c} & \targ{} & \gate{Y_{-c}} & \ctrl{-1} & \qw & \qw
\end{tikzcd}
&
 \begin{tikzcd}[row sep=.1cm, column sep=.1cm]
& \gate[2]{\mathcal{N}_{\mathrm{H}}(\theta_{1})} & \qw &  \qw &\\
& \qw & \gate[2]{\mathcal{N}_{\mathrm{H}}(\theta_{2})} &  \qw &\\
& \gate[2]{\mathcal{N}_{\mathrm{H}}(\theta_{1})} & \qw &  \qw &\\
& \qw & \qw  & \qw &
\end{tikzcd}
& 2\\ \hline

$\mathcal{T}_3^{(+1)}$  &
\begin{tikzcd}[row sep=.1cm, column sep=.1cm]
    \lvert 0\rangle& \gate[1]{X_\pi
    } & \qw & \qw & \targ{}  & \qw & \qw & \qw & \qw & \qw & \qw & \qw & \qw \\
    \lvert 0\rangle& \gate{Y_a} & \gate{X_\pi} & \gate{Y_{-a}}& \ctrl{-1} & \ctrl{1} & \qw & \targ{} & \qw & \qw & \qw & \qw  & \qw \\
    \lvert 0\rangle& \qw & \qw & \qw & \gate{Y_b} & \targ{} & \gate{Y_{-b}} & \ctrl{-1} & \ctrl{1} & \qw & \targ{}  & \qw &\gate{Z_\pi} \\
    \lvert 0\rangle& \qw & \qw & \qw & \qw & \qw & \qw & \gate{Y_c} & \targ{} & \gate{Y_{-c}} & \ctrl{-1} & \qw & \gate{Z_\pi} 
\end{tikzcd}
&
 \begin{tikzcd}[row sep=.1cm, column sep=.1cm]
& \gate[2]{\mathcal{N}_{\mathrm{H}}(\theta_{1})} & \qw &  \qw &\\
& \qw & \gate[2]{\mathcal{N}_{\mathrm{H}}(\theta_{2})} &  \qw &\\
& \gate[2]{\mathcal{N}_{\mathrm{H}}(\theta_{1})} & \qw &  \qw &\\
& \qw & \qw  & \qw &
\end{tikzcd}
& 2 \\ \hline

$\mathcal{T}_1^{(0)}$   &
\begin{tikzcd}[row sep=.1cm, column sep=.1cm]
    \lvert 0\rangle& \gate{H} & \ctrl{1} & \qw  & \qw & \gate{Z_\pi} \\
    \lvert 0\rangle& \qw & \targ{} &  \ctrl{1} & \qw & \gate{X_\pi} \\
    \lvert 0\rangle& \qw & \qw & \targ{}  & \ctrl{1} &  \qw \\
    \lvert 0\rangle& \qw & \qw & \qw  & \targ{} & \gate{X_\pi}
\end{tikzcd}
&
 \begin{tikzcd}[row sep=.1cm, column sep=.1cm]
& \gate[2]{\mathcal{N}_{\mathrm{H}}(\theta_{1})} & \qw &  \qw &\\
& \qw & \gate[2]{\mathcal{N}_{\mathrm{H}}(\theta_{2})} &  \qw &\\
& \gate[2]{\mathcal{N}_{\mathrm{H}}(\theta_{1})} & \qw &  \qw &\\
& \qw & \qw  & \qw &
\end{tikzcd}
& 2 \\ \hline

$\mathcal{T}_2^{(0)}$  &
\begin{tikzcd}[row sep=.1cm, column sep=.1cm]
    \lvert 0\rangle& \gate{H} & \ctrl{1} & \qw  & \qw & \gate{Z_\pi} \\
    \lvert 0\rangle& \qw & \targ{} &  \ctrl{1} & \qw & \gate{X_\pi} \\
    \lvert 0\rangle& \qw & \qw & \targ{}  & \ctrl{1} & \gate{X_\pi} \\
    \lvert 0\rangle& \qw & \qw & \qw  & \targ{} & \qw
\end{tikzcd}
&
 \begin{tikzcd}[row sep=.1cm, column sep=.1cm]
& \gate[2]{\mathcal{N}_{\mathrm{H}}(\theta_{1})} & \qw &  \qw &\\
& \qw & \gate[2]{\mathcal{N}_{\mathrm{H}}(\theta_{2})} &  \qw &\\
& \gate[2]{\mathcal{N}_{\mathrm{H}}(\theta_{1})} & \qw &  \qw &\\
& \qw & \qw  & \qw &
\end{tikzcd}
& 2\\ \hline

$\mathcal{T}_3^{(0)}$  &
\begin{tikzcd}[row sep=.1cm, column sep=.1cm]
    \lvert 0\rangle& \gate{H} & \ctrl{1} & \qw  & \qw & \gate{Z_\pi} \\
    \lvert 0\rangle& \qw & \targ{} &  \ctrl{1} & \qw & \qw \\
    \lvert 0\rangle& \qw & \qw & \targ{}  & \ctrl{1} & \gate{X_\pi} \\
    \lvert 0\rangle& \qw & \qw & \qw  & \targ{} & \gate{X_\pi}
\end{tikzcd}
&
 \begin{tikzcd}[row sep=.1cm, column sep=.1cm]
& \gate[2]{\mathcal{N}_{\mathrm{H}}(\theta_{1})} & \qw &  \qw &\\
& \qw & \gate[2]{\mathcal{N}_{\mathrm{H}}(\theta_{2})} &  \qw &\\
& \gate[2]{\mathcal{N}_{\mathrm{H}}(\theta_{1})} & \qw &  \qw &\\
& \qw & \qw  & \qw &
\end{tikzcd}
& 2\\

\end{tabular}
\label{tab:QinfoN4heis1}
\end{ruledtabular}
\end{table*}

\begin{table*}[htbp]
\centering
\caption{Table S1 continued}
\begin{ruledtabular}
\begin{tabular}{c|ccccccccc}

\textbf{State}& \textbf{Initialization Circuit} & \textbf{Ansatz layer} & \textbf{\# of layers} \\ \hline

$\mathcal{T}_1^{(-1)}$   &
\begin{tikzcd}[row sep=.1cm, column sep=.1cm]
    \lvert 0\rangle& \gate[1]{X_\pi
    } & \qw & \qw & \targ{}  & \qw & \qw & \qw & \qw & \qw & \qw & \qw & \gate{Z_\pi} & \gate{X_\pi} \\
    \lvert 0\rangle& \gate{Y_a} & \gate{X_\pi} & \gate{Y_{-a}}& \ctrl{-1} & \ctrl{1} & \qw & \targ{} & \qw & \qw & \qw & \qw  & \qw & \gate{X_\pi}\\
    \lvert 0\rangle& \qw & \qw & \qw & \gate{Y_b} & \targ{} & \gate{Y_{-b}} & \ctrl{-1} & \ctrl{1} & \qw & \targ{}  & \qw &\gate{Z_\pi} & \gate{X_\pi}\\
    \lvert 0\rangle& \qw & \qw & \qw & \qw & \qw & \qw & \gate{Y_c} & \targ{} & \gate{Y_{-c}} & \ctrl{-1} & \qw & \qw & \gate{X_\pi}
\end{tikzcd}
&
 \begin{tikzcd}[row sep=.1cm, column sep=.1cm]
& \gate[2]{\mathcal{N}_{\mathrm{H}}(\theta_{1})} & \qw &  \qw &\\
& \qw & \gate[2]{\mathcal{N}_{\mathrm{H}}(\theta_{2})} &  \qw &\\
& \gate[2]{\mathcal{N}_{\mathrm{H}}(\theta_{1})} & \qw &  \qw &\\
& \qw & \qw  & \qw &
\end{tikzcd}
& 2\\ \hline

$\mathcal{T}_2^{(-1)}$  &
\begin{tikzcd}[row sep=.1cm, column sep=.1cm]
    \lvert 0\rangle& \gate[1]{X_\pi
    } & \qw & \qw & \targ{}  & \qw & \qw & \qw & \qw & \qw & \qw & \qw & \gate{Z_\pi} & \gate{X_\pi} \\
    \lvert 0\rangle& \gate{Y_a} & \gate{X_\pi} & \gate{Y_{-a}}& \ctrl{-1} & \ctrl{1} & \qw & \targ{} & \qw & \qw & \qw & \qw  & \qw & \gate{X_\pi}\\
    \lvert 0\rangle& \qw & \qw & \qw & \gate{Y_b} & \targ{} & \gate{Y_{-b}} & \ctrl{-1} & \ctrl{1} & \qw & \targ{}  & \qw &\qw & \gate{X_\pi}\\
    \lvert 0\rangle& \qw & \qw & \qw & \qw & \qw & \qw & \gate{Y_c} & \targ{} & \gate{Y_{-c}} & \ctrl{-1} & \qw & \gate{Z_\pi} & \gate{X_\pi}
\end{tikzcd}
&
 \begin{tikzcd}[row sep=.1cm, column sep=.1cm]
& \gate[2]{\mathcal{N}_{\mathrm{H}}(\theta_{1})} & \qw &  \qw &\\
& \qw & \gate[2]{\mathcal{N}_{\mathrm{H}}(\theta_{2})} &  \qw &\\
& \gate[2]{\mathcal{N}_{\mathrm{H}}(\theta_{1})} & \qw &  \qw &\\
& \qw & \qw  & \qw &
\end{tikzcd}
& 2\\ \hline

$\mathcal{T}_3^{(-1)}$  &
\begin{tikzcd}[row sep=.1cm, column sep=.1cm]
    \lvert 0\rangle& \gate[1]{X_\pi
    } & \qw & \qw & \targ{}  & \qw & \qw & \qw & \qw & \qw & \qw & \qw & \gate{Z_\pi} & \gate{X_\pi} \\
    \lvert 0\rangle& \gate{Y_a} & \gate{X_\pi} & \gate{Y_{-a}}& \ctrl{-1} & \ctrl{1} & \qw & \targ{} & \qw & \qw & \qw & \qw  & \gate{Z_\pi} & \gate{X_\pi}\\
    \lvert 0\rangle& \qw & \qw & \qw & \gate{Y_b} & \targ{} & \gate{Y_{-b}} & \ctrl{-1} & \ctrl{1} & \qw & \targ{}  & \qw &\qw & \gate{X_\pi}\\
    \lvert 0\rangle& \qw & \qw & \qw & \qw & \qw & \qw & \gate{Y_c} & \targ{} & \gate{Y_{-c}} & \ctrl{-1} & \qw & \qw & \gate{X_\pi}
\end{tikzcd}
&
 \begin{tikzcd}[row sep=.1cm, column sep=.1cm]
& \gate[2]{\mathcal{N}_{\mathrm{H}}(\theta_{1})} & \qw &  \qw &\\
& \qw & \gate[2]{\mathcal{N}_{\mathrm{H}}(\theta_{2})} &  \qw &\\
& \gate[2]{\mathcal{N}_{\mathrm{H}}(\theta_{1})} & \qw &  \qw &\\
& \qw & \qw  & \qw &
\end{tikzcd}
& 2\\ \hline

$\mathcal{Q}^{(+2)}$  &
\begin{tikzcd}[row sep=.1cm, column sep=.1cm]
    \lvert 0\rangle& \gate{I} & \qw    \\
    \lvert 0\rangle& \gate{I} & \qw    \\
    \lvert 0\rangle& \gate{I} &  \qw   \\
    \lvert 0\rangle& \gate{I} & \qw 
\end{tikzcd}
&
 \begin{tikzcd}[row sep=.1cm, column sep=.1cm]
& \gate[2]{\mathcal{N}_{\mathrm{H}}(\theta_{1})} & \qw &  \qw &\\
& \qw & \gate[2]{\mathcal{N}_{\mathrm{H}}(\theta_{2})} &  \qw &\\
& \gate[2]{\mathcal{N}_{\mathrm{H}}(\theta_{1})} & \qw &  \qw &\\
& \qw & \qw  & \qw &
\end{tikzcd}
& 1\\ \hline

$\mathcal{Q}^{(+1)}$  &
\begin{tikzcd}[row sep=.1cm, column sep=.1cm]
    \lvert 0\rangle& \gate{I} & \qw    \\
    \lvert 0\rangle& \gate{I} & \qw    \\
    \lvert 0\rangle& \gate{I} &  \qw   \\
    \lvert 0\rangle& \gate{X_\pi} & \qw 
\end{tikzcd}
&
 \begin{tikzcd}[row sep=.1cm, column sep=.1cm]
& \gate[2]{\mathcal{N}_{\mathrm{H}}(\theta_{1})} & \qw & \gate{Z(\theta_{3})} & \qw \\
& \qw & \gate[2]{\mathcal{N}_{\mathrm{H}}(\theta_{2})} & \gate{Z(\theta_{4})} & \qw \\
& \gate[2]{\mathcal{N}_{\mathrm{H}}(\theta_{1})} & \qw & \gate{Z(-\theta_{4})} & \qw \\
& \qw & \qw & \gate{Z(-\theta_{3})} & \qw
\end{tikzcd}
& 2\\ \hline

$\mathcal{Q}^{(0)}$  &
\begin{tikzcd}[row sep=.1cm, column sep=.1cm]
    \lvert 0\rangle& \gate{I} & \qw    \\
    \lvert 0\rangle& \gate{X_\pi} & \qw    \\
    \lvert 0\rangle& \gate{I} &  \qw   \\
    \lvert 0\rangle& \gate{X_\pi} & \qw 
\end{tikzcd}
& \begin{tikzcd}[row sep=.1cm, column sep=.1cm]
& \gate[2]{\mathcal{N}_{\mathrm{H}}(\theta_{1})} & \qw & \gate{Z(\theta_{3})} & \qw \\
& \qw & \gate[2]{\mathcal{N}_{\mathrm{H}}(\theta_{2})} & \gate{Z(\theta_{4})} & \qw \\
& \gate[2]{\mathcal{N}_{\mathrm{H}}(\theta_{1})} & \qw & \gate{Z(-\theta_{4})} & \qw \\
& \qw & \qw & \gate{Z(-\theta_{3})} & \qw
\end{tikzcd}
& 2\\ \hline

$\mathcal{Q}^{(-1)}$  &        
\begin{tikzcd}[row sep=.1cm, column sep=.1cm]
    \lvert 0\rangle& \gate{I} & \qw    \\
    \lvert 0\rangle& \gate{X_\pi} & \qw    \\
    \lvert 0\rangle& \gate{X_\pi} &  \qw   \\
    \lvert 0\rangle& \gate{X_\pi} & \qw 
\end{tikzcd}
& \begin{tikzcd}[row sep=.1cm, column sep=.1cm]
& \gate[2]{\mathcal{N}_{\mathrm{H}}(\theta_{1})} & \qw & \gate{Z(\theta_{3})} & \qw \\
& \qw & \gate[2]{\mathcal{N}_{\mathrm{H}}(\theta_{2})} & \gate{Z(\theta_{4})} & \qw \\
& \gate[2]{\mathcal{N}_{\mathrm{H}}(\theta_{1})} & \qw & \gate{Z(-\theta_{4})} & \qw \\
& \qw & \qw & \gate{Z(-\theta_{3})} & \qw
\end{tikzcd}
& 2\\ \hline

$\mathcal{Q}^{(-2)}$  &
\begin{tikzcd}[row sep=.1cm, column sep=.1cm]
    \lvert 0\rangle& \gate{X_\pi} & \qw    \\
    \lvert 0\rangle& \gate{X_\pi} & \qw    \\
    \lvert 0\rangle& \gate{X_\pi} &  \qw   \\
    \lvert 0\rangle& \gate{X_\pi} & \qw 
\end{tikzcd}
&
 \begin{tikzcd}[row sep=.1cm, column sep=.1cm]
& \gate[2]{\mathcal{N}_{\mathrm{H}}(\theta_{1})} & \qw &  \qw &\\
& \qw & \gate[2]{\mathcal{N}_{\mathrm{H}}(\theta_{2})} &  \qw &\\
& \gate[2]{\mathcal{N}_{\mathrm{H}}(\theta_{1})} & \qw &  \qw &\\
& \qw & \qw  & \qw &
\end{tikzcd}
& 1\\

\end{tabular}
\label{tab:QinfoN4heis2}
\end{ruledtabular}
\end{table*}

\begin{table*}[htbp]
\centering
\caption{Circuits for the 8-qubit Heisenberg model}
\begin{ruledtabular}
\begin{tabular}{c|ccccccccc}
\textbf{State}& \textbf{Initialization Circuit} & \textbf{Ansatz layer} & \textbf{\# of layers} \\ \hline

$\mathcal{S}_1$ & 
 \begin{tikzcd}[row sep=.1cm, column sep=.35cm]
\lvert 0\rangle& \gate[1]{X} & \gate[1]{H} & \ctrl{1} & \qw  \\
\lvert 0\rangle& \gate[1]{X} & \qw & \targ{} & \qw & \\
\lvert 0\rangle& \gate[1]{X} & \gate[1]{H} & \ctrl{1} & \qw  \\
\lvert 0\rangle& \gate[1]{X} & \qw & \targ{} & \qw \\
\lvert 0\rangle& \gate[1]{X} & \gate[1]{H} & \ctrl{1} & \qw  \\
\lvert 0\rangle& \gate[1]{X} & \qw & \targ{} & \qw \\
\lvert 0\rangle& \gate[1]{X} & \gate[1]{H} & \ctrl{1} & \qw  \\
\lvert 0\rangle& \gate[1]{X} & \qw & \targ{} & \qw 
\end{tikzcd}
&
 \begin{tikzcd}[row sep=.1cm, column sep=.35cm]
&\gate[2]{\mathcal{N}_{\mathrm{H}}(\theta_{1})} & \qw  & \qw \\
&\qw & \gate[2]{\mathcal{N}_{\mathrm{H}}(\theta_{3})}  & \qw \\
&\gate[2]{\mathcal{N}_{\mathrm{H}}(\theta_{2})} & \qw  & \qw \\
&\qw & \gate[2]{\mathcal{N}_{\mathrm{H}}(\theta_{4})} & \qw \\
&\gate[2]{\mathcal{N}_{\mathrm{H}}(\theta_{2})} & \qw  & \qw \\
&\qw & \gate[2]{\mathcal{N}_{\mathrm{H}}(\theta_{3})} & \qw \\
&\gate[2]{\mathcal{N}_{\mathrm{H}}(\theta_{1})} & \qw  & \qw \\
&\qw & \qw & \qw
\end{tikzcd}
& 3 \\ \hline

$\mathcal{T}_1^{(0)}$   &
\begin{tikzcd}[row sep=.1cm, column sep=.35cm]
\lvert 0\rangle& \gate[1]{X} & \gate[1]{H} & \ctrl{1} & \qw  \\
\lvert 0\rangle& \gate[1]{X} & \qw & \targ{} & \qw & \\
\lvert 0\rangle& \gate[1]{X} & \gate[1]{H} & \ctrl{1} & \qw  \\
\lvert 0\rangle& \gate[1]{X} & \qw & \targ{} & \qw \\
\lvert 0\rangle& \qw & \gate[1]{H} & \ctrl{1} & \qw  \\
\lvert 0\rangle& \gate[1]{X} & \qw & \targ{} & \qw \\
\lvert 0\rangle& \gate[1]{X} & \gate[1]{H} & \ctrl{1} & \qw  \\
\lvert 0\rangle& \gate[1]{X} & \qw & \targ{} & \qw 
\end{tikzcd}
&
 \begin{tikzcd}[row sep=.1cm, column sep=.35cm]
& \gate[2]{\mathcal{N}_{\mathrm{H}}(\theta_{1})} & \qw  & \qw \\
& \qw & \gate[2]{\mathcal{N}_{\mathrm{H}}(\theta_{3})}  & \qw \\
& \gate[2]{\mathcal{N}_{\mathrm{H}}(\theta_{2})} & \qw  & \qw \\
& \qw & \gate[2]{\mathcal{N}_{\mathrm{H}}(\theta_{4})} & \qw \\
& \gate[2]{\mathcal{N}_{\mathrm{H}}(-\theta_{2})} & \qw  & \qw \\
& \qw & \gate[2]{\mathcal{N}_{\mathrm{H}}(-\theta_{3})} & \qw \\
& \gate[2]{\mathcal{N}_{\mathrm{H}}(-\theta_{1})} & \qw  & \qw \\
& \qw & \qw & \qw
\end{tikzcd}
& 4 \\ \hline

$\mathcal{T}_2^{(0)}$  &
\begin{tikzcd}[row sep=.1cm, column sep=.35cm]
\lvert 0\rangle& \gate[1]{X} & \gate[1]{H} & \ctrl{1} & \qw  \\
\lvert 0\rangle& \gate[1]{X} & \qw & \targ{} & \qw & \\
\lvert 0\rangle& \gate[1]{X} & \gate[1]{H} & \ctrl{1} & \qw  \\
\lvert 0\rangle& \gate[1]{X} & \qw & \targ{} & \qw \\
\lvert 0\rangle& \qw & \gate[1]{H} & \ctrl{1} & \qw  \\
\lvert 0\rangle& \gate[1]{X} & \qw & \targ{} & \qw \\
\lvert 0\rangle& \gate[1]{X} & \gate[1]{H} & \ctrl{1} & \qw  \\
\lvert 0\rangle& \gate[1]{X} & \qw & \targ{} & \qw 
\end{tikzcd}
,
\begin{tikzcd}[row sep=.1cm, column sep=.35cm]
\lvert 0\rangle& \gate[1]{X} & \gate[1]{H} & \ctrl{1} & \qw  \\
\lvert 0\rangle& \gate[1]{X} & \qw & \targ{} & \qw & \\
\lvert 0\rangle& \qw & \gate[1]{H} & \ctrl{1} & \qw  \\
\lvert 0\rangle& \gate[1]{X} & \qw & \targ{} & \qw \\
\lvert 0\rangle& \gate[1]{X} & \gate[1]{H} & \ctrl{1} & \qw  \\
\lvert 0\rangle& \gate[1]{X} & \qw & \targ{} & \qw \\
\lvert 0\rangle& \gate[1]{X} & \gate[1]{H} & \ctrl{1} & \qw  \\
\lvert 0\rangle& \gate[1]{X} & \qw & \targ{} & \qw 
\end{tikzcd}
&
 \begin{tikzcd}[row sep=.1cm, column sep=.35cm]
& \gate[2]{\mathcal{N}_{\mathrm{H}}(\theta_{1})} & \qw  & \qw \\
& \qw & \gate[2]{\mathcal{N}_{\mathrm{H}}(\theta_{3})}  & \qw \\
& \gate[2]{\mathcal{N}_{\mathrm{H}}(\theta_{2})} & \qw  & \qw \\
& \qw & \gate[2]{\mathcal{N}_{\mathrm{H}}(\theta_{4})} & \qw \\
& \gate[2]{\mathcal{N}_{\mathrm{H}}(-\theta_{2})} & \qw  & \qw \\
& \qw & \gate[2]{\mathcal{N}_{\mathrm{H}}(-\theta_{3})} & \qw \\
& \gate[2]{\mathcal{N}_{\mathrm{H}}(-\theta_{1})} & \qw  & \qw \\
& \qw & \qw & \qw
\end{tikzcd}
& 4\\

\end{tabular}
\label{tab:QinfoN8heis}
\end{ruledtabular}
\end{table*}

\begin{table*}[htbp]
\centering
\caption{Circuits for the 4-qubit transverse Ising model}
\begin{ruledtabular}
\begin{tabular}{c|ccccccccc}
\textbf{State}& \textbf{Initialization Circuit} & \textbf{Ansatz layer} & \textbf{\# of layers} \\ \hline

$E_1$ & 
 \begin{tikzcd}[row sep=.1cm, column sep=.35cm]
\lvert 0\rangle& \qw  & \qw & \qw \\
\lvert 0\rangle& \gate[1]{H} & \ctrl{1} & \qw \\
\lvert 0\rangle& \qw & \targ{} & \qw \\
\lvert 0\rangle& \qw & \qw & \qw
\end{tikzcd}
&
 \begin{tikzcd}[row sep=.1cm, column sep=.35cm]
& \gate[2]{\mathcal{N}_{\mathrm{I}}(\theta_{1})} & \qw & \gate{Z(\theta_{3})} & \qw \\
& \qw & \gate[2]{\mathcal{N}_{\mathrm{I}}(\theta_{2})} & \gate{Z(\theta_{4})} & \qw \\
& \gate[2]{\mathcal{N}_{\mathrm{I}}(\theta_{1})} & \qw & \gate{Z(\theta_{4})} & \qw \\
& \qw & \qw & \gate{Z(\theta_{3})} & \qw
\end{tikzcd}
& 2 \\ \hline

$E_2$   &
\begin{tikzcd}[row sep=.1cm, column sep=.35cm]
\lvert 0\rangle& \qw  & \qw & \qw & \qw \\
\lvert 0\rangle& \gate[1]{H} & \ctrl{1} & \gate{Z} & \qw \\
\lvert 0\rangle& \gate[1]{X} & \targ{} & \qw & \qw\\
\lvert 0\rangle& \qw & \qw & \qw & \qw 
\end{tikzcd}
&
 \begin{tikzcd}[row sep=.1cm, column sep=.35cm]
& \gate[2]{\mathcal{N}_{\mathrm{I}}(\theta_{1})} & \qw & \gate{Z(\theta_{3})} & \qw \\
& \qw & \gate[2]{\mathcal{N}_{\mathrm{I}}(\theta_{2})} & \gate{Z(\theta_{4})} & \qw \\
& \gate[2]{\mathcal{N}_{\mathrm{I}}(\theta_{1})} & \qw & \gate{Z(\theta_{4})} & \qw \\
& \qw & \qw & \gate{Z(\theta_{3})} & \qw
\end{tikzcd}
& 2 \\ \hline

$E_3$  &
\begin{tikzcd}[row sep=.1cm, column sep=.35cm]
\lvert 0\rangle& \qw  & \qw & \qw  \\
\lvert 0\rangle& \gate[1]{H} & \ctrl{1} & \qw \\
\lvert 0\rangle& \gate[1]{X} & \targ{} & \qw\\
\lvert 0\rangle& \qw & \qw & \qw  
\end{tikzcd}
&
 \begin{tikzcd}[row sep=.1cm, column sep=.35cm]
& \gate[2]{\mathcal{N}_{\mathrm{I}}(\theta_{1})} & \qw & \gate{Z(\theta_{3})} & \qw \\
& \qw & \gate[2]{\mathcal{N}_{\mathrm{I}}(\theta_{2})} & \gate{Z(\theta_{4})} & \qw \\
& \gate[2]{\mathcal{N}_{\mathrm{I}}(\theta_{1})} & \qw & \gate{Z(\theta_{4})} & \qw \\
& \qw & \qw & \gate{Z(\theta_{3})} & \qw
\end{tikzcd}
& 2\\ 

\end{tabular}
\label{tab:QinfoN4Ising}
\end{ruledtabular}
\end{table*}

\begin{table*}[htbp]
\centering
\caption{Circuits for the 8-qubit transverse Ising model}
\begin{ruledtabular}
\begin{tabular}{c|ccccccccc}
\textbf{State}& \textbf{Initialization Circuit} & \textbf{Ansatz layer} & \textbf{\# of layers} \\ \hline

$E_1$ & 
 \begin{tikzcd}[row sep=.1cm, column sep=.35cm]
\lvert 0\rangle& \qw & \qw & \qw \\
\lvert 0\rangle& \qw & \qw & \qw \\
\lvert 0\rangle& \qw  & \qw & \qw \\
\lvert 0\rangle& \gate[1]{H} & \ctrl{1} & \qw \\
\lvert 0\rangle& \qw & \targ{} & \qw \\
\lvert 0\rangle& \qw & \qw & \qw \\
\lvert 0\rangle& \qw & \qw & \qw \\
\lvert 0\rangle& \qw & \qw & \qw
\end{tikzcd}
&
 \begin{tikzcd}[row sep=.1cm, column sep=.35cm]
& \gate[2]{\mathcal{N}_{\mathrm{I}}(\theta_{1})} & \qw & \gate{Z(\theta_{5})} & \qw \\
& \qw & \gate[2]{\mathcal{N}_{\mathrm{I}}(\theta_{3})} & \gate{Z(\theta_{6})} & \qw \\
& \gate[2]{\mathcal{N}_{\mathrm{I}}(\theta_{2})} & \qw & \gate{Z(\theta_{7})} & \qw \\
& \qw & \gate[2]{\mathcal{N}_{\mathrm{I}}(\theta_{4})} & \gate{Z(\theta_{8})} & \qw \\
& \gate[2]{\mathcal{N}_{\mathrm{I}}(\theta_{2})} & \qw & \gate{Z(\theta_{8})} & \qw \\
& \qw & \gate[2]{\mathcal{N}_{\mathrm{I}}(\theta_{3})} & \gate{Z(\theta_{7})} & \qw \\
& \gate[2]{\mathcal{N}_{\mathrm{I}}(\theta_{1})} & \qw & \gate{Z(\theta_{6})} & \qw \\
& \qw & \qw & \gate{Z(\theta_{5})} & \qw 
\end{tikzcd}
& 3 \\ \hline

$E_2$   &
 \begin{tikzcd}[row sep=.1cm, column sep=.35cm]
\lvert 0\rangle& \qw & \qw & \qw& \qw \\
\lvert 0\rangle& \qw & \qw & \qw& \qw \\
\lvert 0\rangle& \qw  & \qw & \qw& \qw \\
\lvert 0\rangle& \gate[1]{H} & \ctrl{1} & \gate[1]{Z}& \qw \\
\lvert 0\rangle& \gate[1]{X} & \targ{} & \qw& \qw \\
\lvert 0\rangle& \qw & \qw & \qw& \qw \\
\lvert 0\rangle& \qw & \qw & \qw& \qw \\
\lvert 0\rangle& \qw & \qw & \qw& \qw
\end{tikzcd}
&
 \begin{tikzcd}[row sep=.1cm, column sep=.35cm]
& \gate[2]{\mathcal{N}_{\mathrm{I}}(\theta_{1})} & \qw & \gate{Z(\theta_{5})} & \qw \\
& \qw & \gate[2]{\mathcal{N}_{\mathrm{I}}(\theta_{3})} & \gate{Z(\theta_{6})} & \qw \\
& \gate[2]{\mathcal{N}_{\mathrm{I}}(\theta_{2})} & \qw & \gate{Z(\theta_{7})} & \qw \\
& \qw & \gate[2]{\mathcal{N}_{\mathrm{I}}(\theta_{4})} & \gate{Z(\theta_{8})} & \qw \\
& \gate[2]{\mathcal{N}_{\mathrm{I}}(\theta_{2})} & \qw & \gate{Z(\theta_{8})} & \qw \\
& \qw & \gate[2]{\mathcal{N}_{\mathrm{I}}(\theta_{3})} & \gate{Z(\theta_{7})} & \qw \\
& \gate[2]{\mathcal{N}_{\mathrm{I}}(\theta_{1})} & \qw & \gate{Z(\theta_{6})} & \qw \\
& \qw & \qw & \gate{Z(\theta_{5})} & \qw 
\end{tikzcd}
& 3 \\ \hline

$E_3$  &
 \begin{tikzcd}[row sep=.1cm, column sep=.35cm]
\lvert 0\rangle& \qw & \qw & \qw \\
\lvert 0\rangle& \qw & \qw & \qw \\
\lvert 0\rangle& \qw  & \qw & \qw \\
\lvert 0\rangle& \gate[1]{H} & \ctrl{1} & \qw \\
\lvert 0\rangle& \gate[1]{X} & \targ{} & \qw \\
\lvert 0\rangle& \qw & \qw & \qw \\
\lvert 0\rangle& \qw & \qw & \qw \\
\lvert 0\rangle& \qw & \qw & \qw
\end{tikzcd}
&
 \begin{tikzcd}[row sep=.1cm, column sep=.35cm]
& \gate[2]{\mathcal{N}_{\mathrm{I}}(\theta_{1})} & \qw & \gate{Z(\theta_{5})} & \qw \\
& \qw & \gate[2]{\mathcal{N}_{\mathrm{I}}(\theta_{3})} & \gate{Z(\theta_{6})} & \qw \\
& \gate[2]{\mathcal{N}_{\mathrm{I}}(\theta_{2})} & \qw & \gate{Z(\theta_{7})} & \qw \\
& \qw & \gate[2]{\mathcal{N}_{\mathrm{I}}(\theta_{4})} & \gate{Z(\theta_{8})} & \qw \\
& \gate[2]{\mathcal{N}_{\mathrm{I}}(\theta_{2})} & \qw & \gate{Z(\theta_{8})} & \qw \\
& \qw & \gate[2]{\mathcal{N}_{\mathrm{I}}(\theta_{3})} & \gate{Z(\theta_{7})} & \qw \\
& \gate[2]{\mathcal{N}_{\mathrm{I}}(\theta_{1})} & \qw & \gate{Z(\theta_{6})} & \qw \\
& \qw & \qw & \gate{Z(\theta_{5})} & \qw 
\end{tikzcd}
& 3\\ 

\end{tabular}
\label{tab:QinfoN8Ising}
\end{ruledtabular}
\end{table*}

\section{Device and measurement setup}
\begin{figure*}
  \centering
  \includegraphics[width=0.85\textwidth]{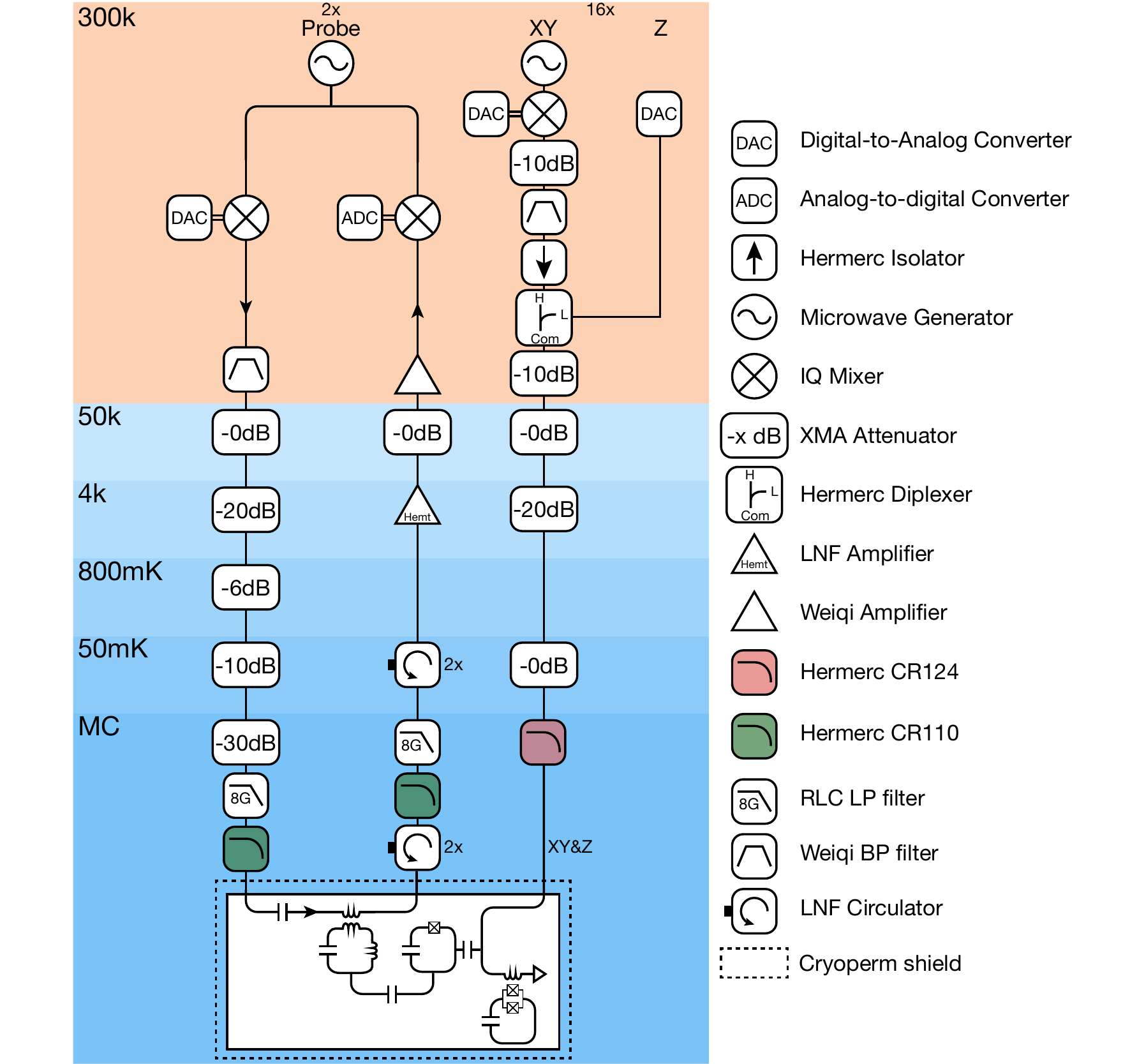}
  \caption[]{\textbf{Schematic diagram of the measurement setup.}
  One set of the readout resonator, qubit, tunable coupler, the shared control line, and corresponding wiring is shown.}
  \label{Fig_Device}
\end{figure*}

The processor is made of aluminum on sapphire using a recipe similar to that in \cite{chu2023scalable} and bonded in an aluminum sample box which is then mounted inside a Blufors LD400 dilution refrigerator with a base temperature of about 10 mK. The sample box is shielded by two layers of cryo-perm cylinders. Figure \ref{Fig_Device} depicts the schematic diagram of the measurement setup. 

The XY microwave signals used for single-qubit operations are generated using the IQ mixing. A bandpass filter at the qubit frequency and an isolator are used for suppressing the spurious and reflection. Fast Z pulses used for two-qubit gates are combined with the XY pulses at room temperature using a diplexer, and then attenuated and filtered at multiple temperature stages for further noise reduction. A customized low-pass IR filter is added to the control lines at the mixing-chamber stage which attenuates signals at the qubit frequency by approximately 25 dB while allowing low-frequency signals to pass through, serving as an equalizer. 

The multiplexed readout signals are attenuated and filtered before being fed into the readout transmission lines. The output signals are then sequentially amplified by a HEMT amplifier at 4~K and a room temperature amplifier before down-conversion and acquisition by the digitizer. 

\begin{figure}
  \centering
  \includegraphics[width=0.6\textwidth]{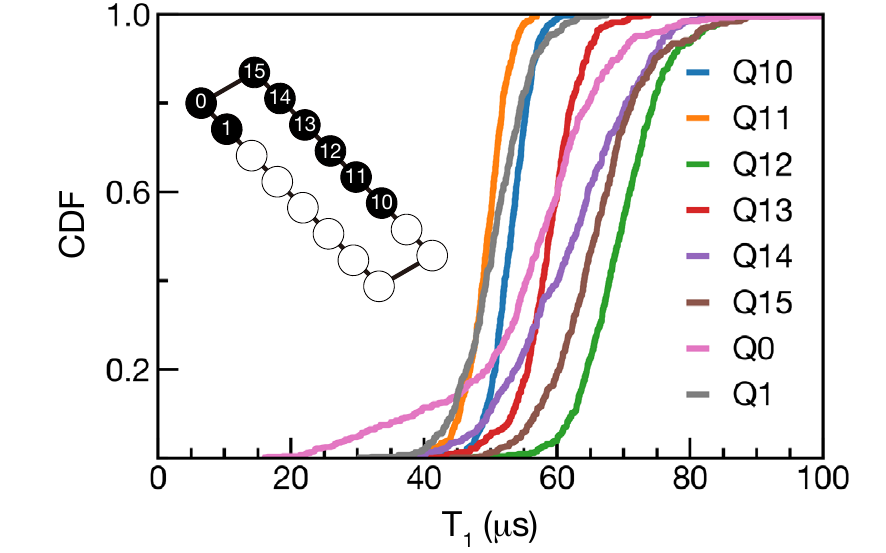}
  \caption[]{\textbf{Qubit $T_1$ time.} 
  Cumulative distribution of the repeated measured $T_1$ relaxation times for the selected eight qubits (inset).  }
  \label{Fig_T1_distribution}
\end{figure}

Repeated characterization result of the $T_1$ relaxation time is shown in Fig.~\ref{Fig_T1_distribution}. The $T_1$ times of the eight selected qubits used in our experiment generally ranges from 40~$\mu$s to 80~$\mu$s, each showing a typical fluctuation of 10-20~$\mu$s. 
The average flux fluctuation in our device is estimated to be approximately 60~$\mu\Phi_0$. The measured ZZ interaction between neighboring qubits is around 10 kHz and therefore considered negligible.
More device details are listed in Table \ref{table:Device_paras}. 

\begin{table*}[htbp]
\centering
\begin{threeparttable}

\begin{tabular}{ |m{4.5cm} | m{1.3cm}| m{1.3cm} |m{1.3cm}| m{1.3cm} |m{1.3cm}| m{1.3cm} | m{1.3cm}| m{1.3cm}| }
    \hline \hline
    Qubit & $Q_{10}^*$ & $Q_{11}^*$ & $Q_{12}^*$ & $Q_{13}^*$ & $Q_{14}$ & $Q_{15}$ & $Q_{0}$ & $Q_{1}$      \\ \hline
    Cavity frequency (GHz)&6.928&7.093&6.968&7.122& 6.998& 7.172& 6.895& 7.052   \\
    Qubit frequency (GHz)&3.704& 4.221& 3.799& 4.219& 3.705& 4.235& 3.780 & 4.223   \\
    Anharmonicity (MHz)  &-213.0&-203.4&-211.5&-202.8&-212.8&-204.5&-211.2&-201.4   \\
    Dispersive shift of $|1\rangle$ (MHz)  &0.6&0.6&0.6&0.45&1&0.5&0.6&0.5 \\
    1-qubit gate error (simul.)(\%)&0.21&0.63&0.53&0.32&0.15&0.95&0.25&0.89  \\
    Lifetime, $T_\mathrm{1}$ ($\mu s$) & 53 & 49 & 69 & 58 & 61 & 65 & 54 & 50 \\
    Ramsey decay time, $T_{2}^*$ ($\mu s$) & 8 & 22 & 7 & 6 & 7 & 11 & 12 & 15   \\
    Echo decay time, $T_{\mathrm{2E}}$ ($\mu s$) & 20 & 29 & 20 & 19 & 24 & 22 & 30 & 19   \\
    Readout fidelity of $\lvert 0\rangle (\%)$&0.927&0.940&0.901&0.926&0.861&0.942&0.924&0.944\\
    Readout fidelity of $\lvert 1\rangle^{**} (\%) $&0.886&0.925&0.887&0.879&0.843&0.888&0.868&0.896\\
    \hline
    Coupler (Group A) & \multicolumn{2}{c}{$C_{10-11}$} &  \multicolumn{2}{|c|}{$C_{12-13}$}  & \multicolumn{2}{c}{$C_{14-15}$} &  \multicolumn{2}{|c|}{$C_{00-01}$} \\ \hline
    
    CZ gate error  (simul.)(\%)   & \multicolumn{2}{c|}{0.96}          & \multicolumn{2}{c|}{0.69} & \multicolumn{2}{c|}{1.24}&\multicolumn{2}{c|}{1.33} \\ 
    
    $ZZ^{***}$ $\mathrm{(kHz)}$ & \multicolumn{2}{c|}{10.4}&\multicolumn{2}{c|}{12.2}&\multicolumn{2}{c|}{15.3}&\multicolumn{2}{c|}{12.2}\\ \hline

    Coupler (Group B)& &  \multicolumn{2}{c|}{$C_{11-12}$} & \multicolumn{2}{c|}{$C_{13-14}$} & \multicolumn{2}{c|}{$C_{15-00}$}&  \\ \hline
    
    CZ gate error  (simul.)(\%)&&  \multicolumn{2}{c|}{2.12} & \multicolumn{2}{c|}{0.99} & \multicolumn{2}{c|}{1.24} & \\ 
    
    $ZZ^{***}$ $\mathrm{(kHz)}$&& \multicolumn{2}{c|}{10.0}&\multicolumn{2}{c|}{10.5}&\multicolumn{2}{c|}{9.5}& \\ \hline \hline
\end{tabular}

\begin{tablenotes}
    \footnotesize
    \item[*] Qubits used in the 4-qubit experiment. See Fig.~\ref{Fig_T1_distribution} for the indexing information of the qubits. 
    \item[**] Both the $|2\rangle$ and $|1\rangle$ state have been treated as the $|1\rangle$ state with the shelving protocol.
    \item[***] The ZZ interaction strength between neighboring qubits is measured when couplers are biased at their idling point.
\end{tablenotes}

\end{threeparttable}

\caption{Device parameters and gate fidelities.}

\label{table:Device_paras}

\end{table*}

\section{Single-qubit gates}
\begin{figure*}
  \centering
  \includegraphics[width=0.5\textwidth]{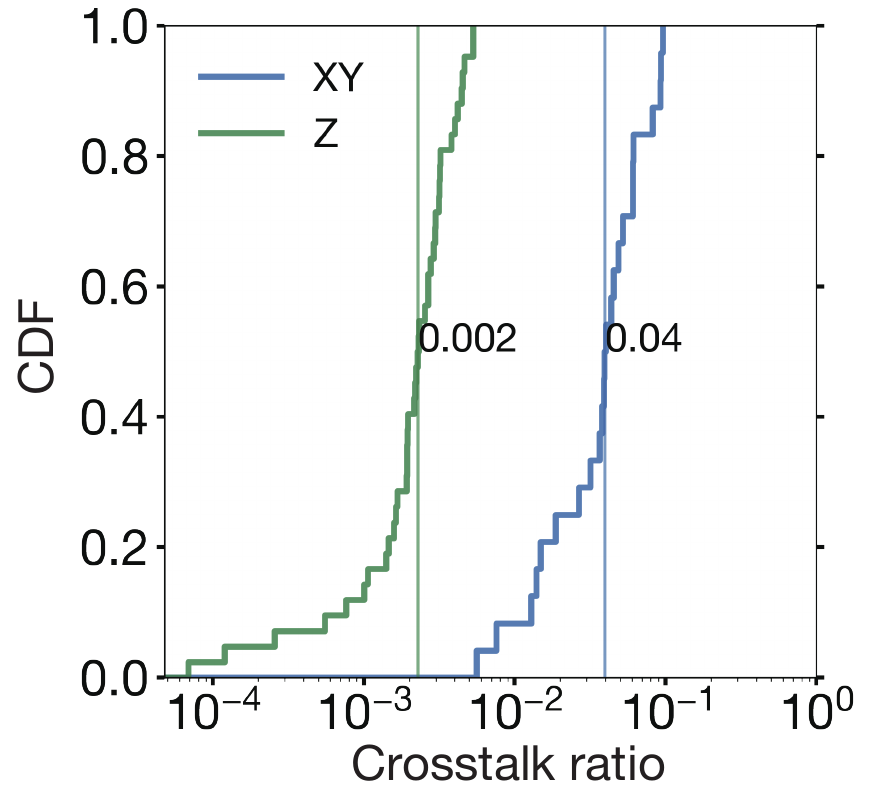}
  \caption[]{\textbf{Crosstalk characterization.}  
  Cumulative distribution of measured microwave (XY) and flux (Z) crosstalk ratios with the median values indicated.}
  \label{Fig_xtalk}
\end{figure*}

The microwave signal or XY crosstalk is a major source of errors when performing parallel single-qubit gates. In our experiment, we use a hybrid strategy to combat the XY crosstalk. For qubits with significant crosstalk and with frequencies close to each other (usually the next-nearest-neighbor qubits), we calibrate and add the compensation signal for crosstalk cancellation~\cite{nuerbolati2022canceling}. Aside from those, we rely on the isomorphous waveform strategy presented in the main text which compiles the circuit into restless single-qubit cycles and automatically corrects for crosstalk. The flux or Z crosstalk can be effectively cancelled by adding compensating signals. The statistics of both XY and Z crosstalk - 4\% for XY and 0.2\% for Z (median value) - in our device are shown in Fig.~\ref{Fig_xtalk}.

For single-qubit gates, we employ the U3 decomposition~\cite{mckay2017efficient} with the need for calibrating $X_{\pi/2}$ pulses only.
We use 30-ns-long $X_{\pi/2}$. We add an additional 10-dB attenuation at room temperature to rescale the pulse amplitudes to approximately half of the maximum DAC output range for better signal quality. To simultaneously calibrate these $X_{\pi/2}$ gates, we use tbe pulse train technique which progressively tunes up the pulse amplitude and DRAG coefficients with increasing cycle number. We characterize our calibrated single-qubit gate errors using simultaneous cross-entropy benchmarking (XEB), with the result shown in Table \ref{table:Device_paras}. 
The simultaneous single-qubit gate fidelities are limited by $T_2$ times and residual crosstalk effect.

\section{Two-qubit gates}
\begin{figure*}
  \centering
  \includegraphics[width=0.6\textwidth]{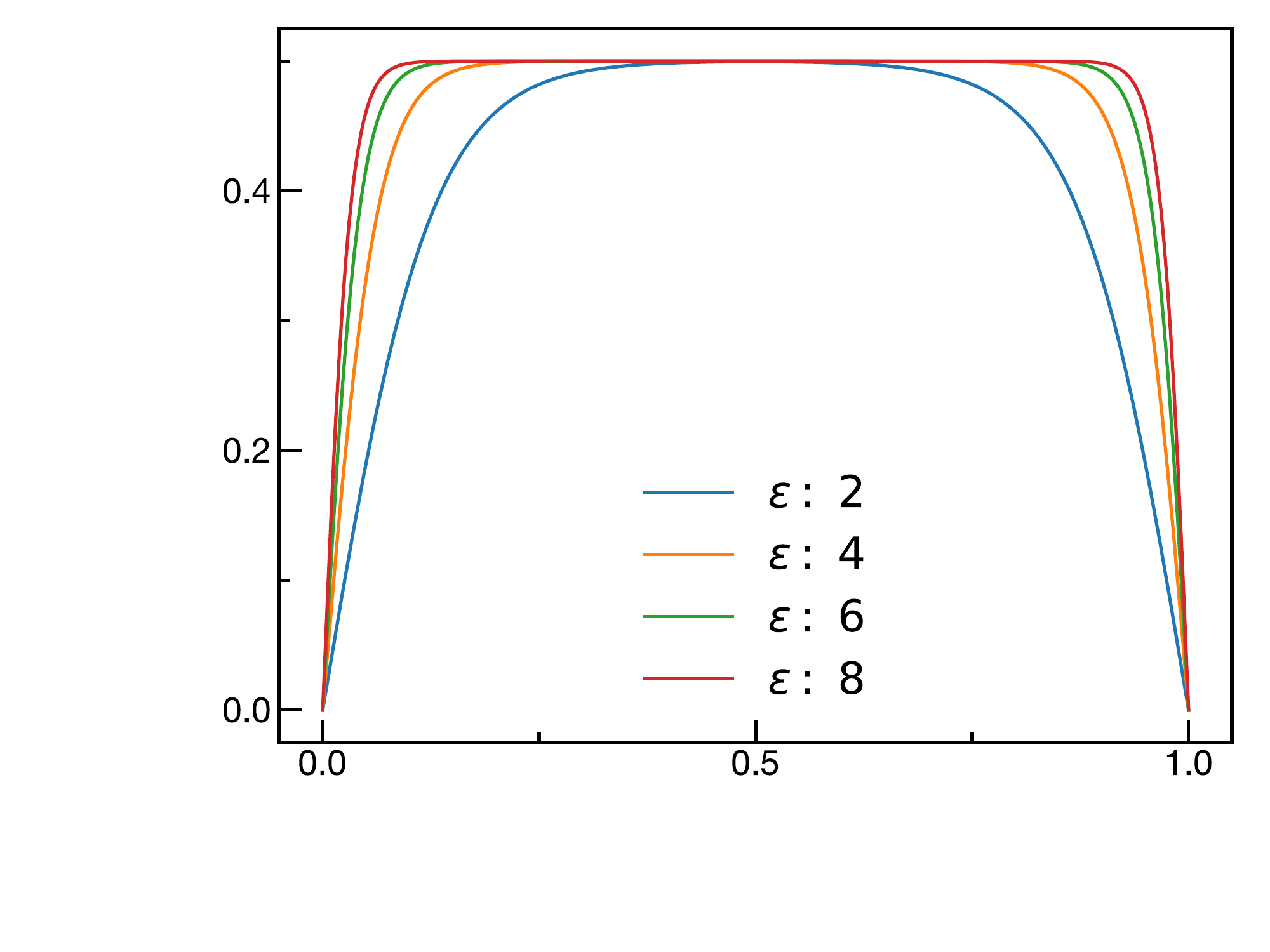}
  \caption[]{\textbf{CZ pulse shape.} 
  The unipolar hyperbolic tangent pulse with different values for the edge parameter $\epsilon$.}
  \label{Fig_pulseshape}
\end{figure*}

The CNOT gates used in our experiment are synthesized from native CZ gates and single-qubit gates. We use the coupler-assisted adiabatic CZ gate scheme demonstrated in our previous work~\cite{xu2020high}. The bipolar net-zero pulse technique is used here for improving robustness against low-frequency flux noise and minimizing signal distortion. For a unipolar half, we employ a hyperbolic tangent pulse shape described by
\begin{equation}
V(t)= A\frac{\tanh(4\epsilon(1/2-|t|/\tau))}{\tanh(2\epsilon)},\;  -\tau/2<t<\tau/2
\label{tanh}
\end{equation}
where $A$ and $\tau$ are the pulse amplitude and width respectively. $\epsilon$ is a parameter controlling pulse shape. Pulse profiles with different $\epsilon$ parameters are shown in Fig.~\ref{Fig_pulseshape}. Larger $\epsilon$ value tends to make the pulse rise more rapidly and hence less adiabatic. In our setup, we set all the CZ pulse duration to 60+60~ns ($\tau=60$~ns for each pole) as a trade-off between leakage and decoherence errors. 

\begin{figure*}
  \centering
  \includegraphics[width=0.85\textwidth]{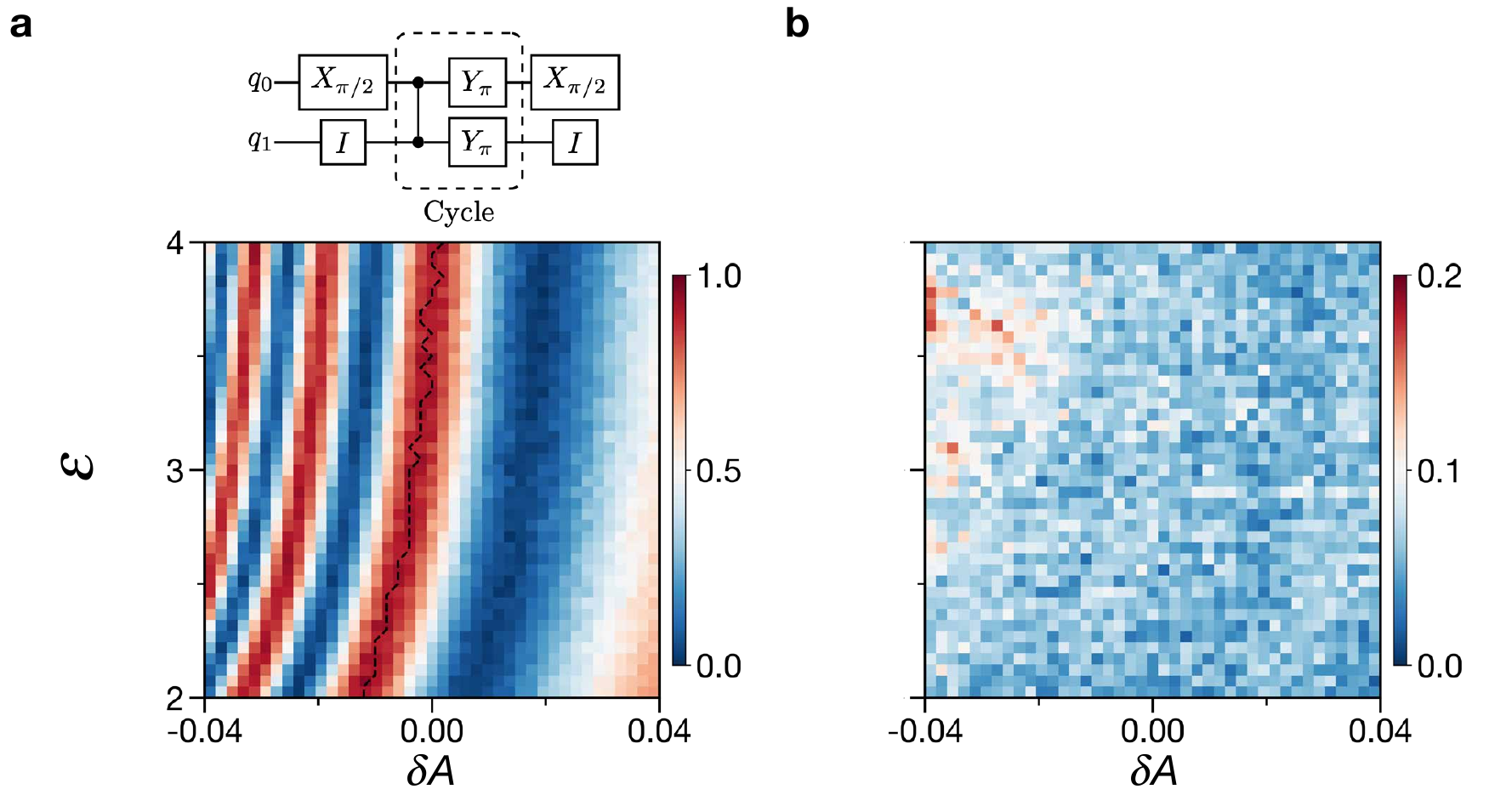}
  \caption[]{\textbf{CZ pulse shape and leakage.} 
  \textbf{a,} The measured populations of the target qubit $q_0$ versus the shape parameter $\epsilon$ and the pulse amplitude $\delta A$ referenced to a certain value under the echoed CZ pulse train sequence shown above. Here the number of repeated cycles is 4. The red fringe in the middle corresponds to a controlled phase of $\pi$. 
  \textbf{b,} The measured populations of the control qubit $q_1$ under the same sequence described in \textbf{a}. The region with more population (red) indicates more leakage.}
  \label{Fig_cali_2d}
\end{figure*}

\begin{figure*}
    \centering
    \includegraphics[width=0.85\textwidth]{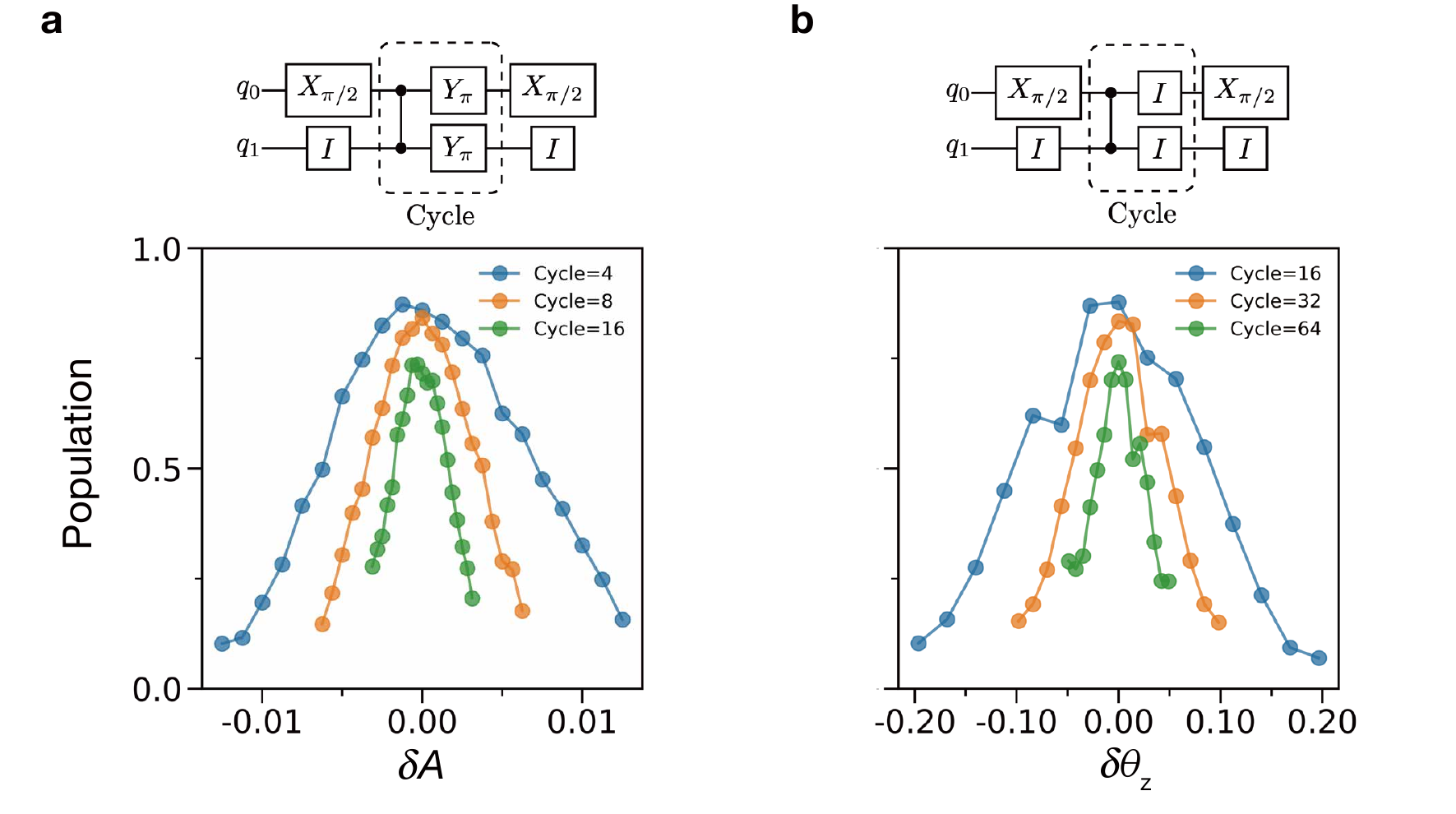}
    \caption[]{\textbf{Controlled phase and local Z phase calibration.} 
    \textbf{a,} Progressive tuning of the pulse amplitude with respect to the controlled phase calibration with increasing number of cycles. On the top is the calibration circuit.
    \textbf{b,} Progressive tuning of the local Z phase correction induced by the CZ pulse with increasing number of cycles. On the top is the calibration circuit.}
    \label{Fig_cali}
  \end{figure*}

During the CZ gate calibration, we first calibrate the flux signal (Z) distortion and use inverse-IIR filters to predistort our pulses. Then, we calibrate the flux (Z) crosstalk among different control lines and cancel it with compensation pulse. The shape parameter $\epsilon$ is decided by performig the echoed CZ pulse train as a function of $\epsilon$ and pulse amplitude $A$ (Fig.~\ref{Fig_cali_2d}\textbf{(a-b)}). The control qubit population reveals the information about the controlled phase while the target qubit population reveals the information about the leakage. Monitoring both helps us determine the approriate value for $\epsilon$. Here we choose $\epsilon\approx 2$ in our experiment.

Once the shape of the CZ pulse is determined, the only left pulse parameter to calibrate is the pulse amplitude $A$, which can be progressively tuned using the same echoed pulse train sequence (Fig.~\ref{Fig_cali}\textbf{(a)}). 
A similar procedure is followed to obtain the local Z phase correction for both qubits (Fig.~\ref{Fig_cali}\textbf{(b)}). 
Note that we devide the CZ gates in the 8-qubit chain into two groups (Table \ref{table:Device_paras}). These calibrations are simultaneously performed for qubit pairs in a same group.
The calibrated CZ gates are simultaneously benchmarked using parallel two-qubit XEB. The average gate fidelities are approximately 0.99 after subtracting the single-qubit gate errors (Table \ref{table:Device_paras}).

\section{Readout} 
To improve the fidelity of the readout, we employ the shelving technique as described in Ref.~\cite{elder2020high}. This method involves the addition of extra $\pi_{ef}$ pulses to all qubits before executing the measurement pulse. When the demodulated signal falls within the $|e\rangle$ or $|f\rangle$ region, it is regarded as the $|e\rangle$ state. Consequently, any finite relaxation from the $|f\rangle$ to the $|e\rangle$ state during the readout pulse does not contribute to single-shot readout errors. The $\pi_{ef}$ pulses are calibrated concurrently, with a pulse width of 200 ns. The readout correction matrix is determined by preparing all the $2^N$ computational states and measuring the probability of each resulting outcome.

\section{Potential landscape}

\begin{figure*}
  \centering
  \includegraphics[width=0.95\textwidth]{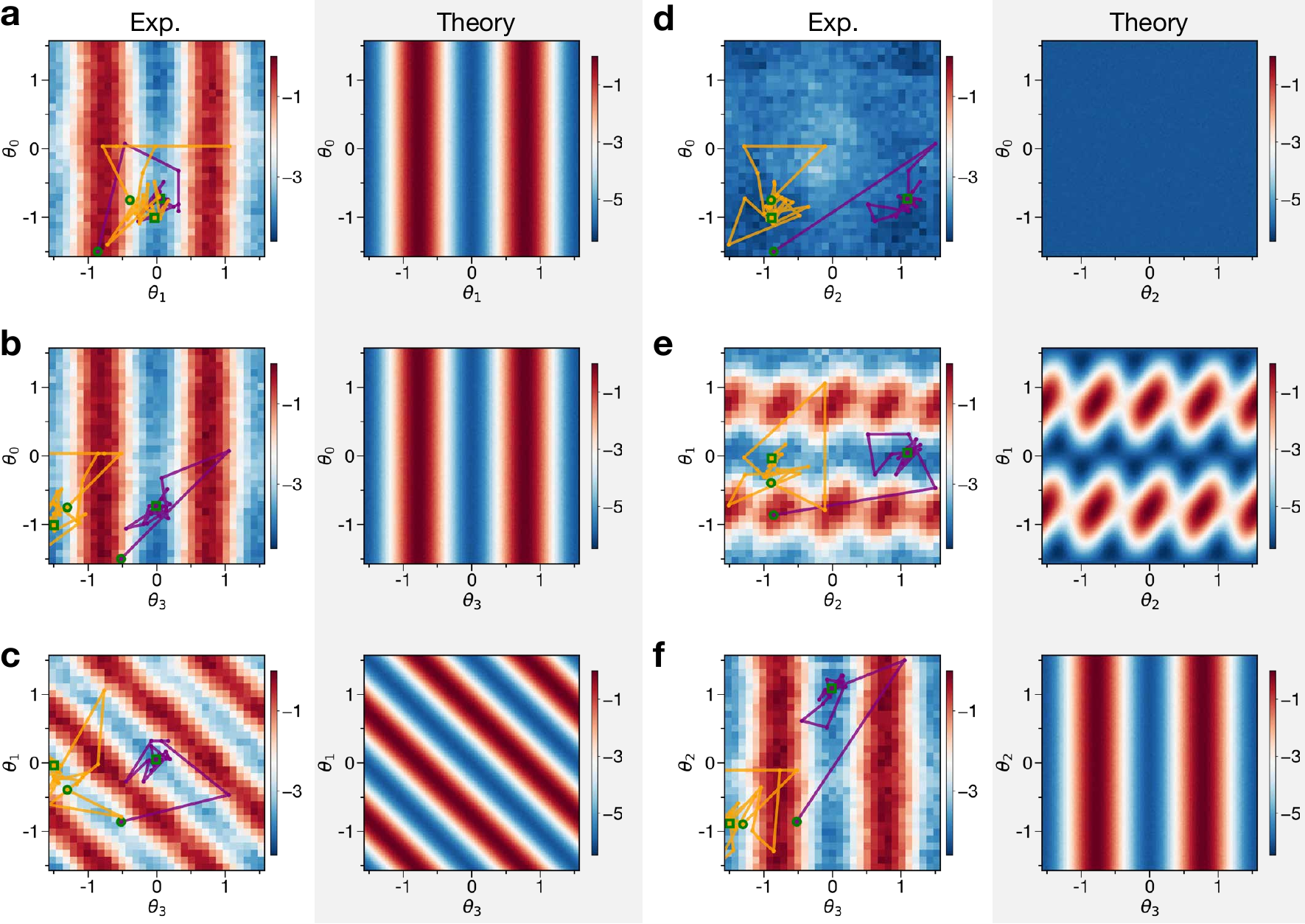}
  \caption[]{\textbf{Energy landscape of the $\mathcal{S}_1$ state in the 4-qubit Heisenberg model.} 
  \textbf{a-f,} The measured and theoretical values of the cost function in the $\mathcal{S}_1$ case as a function of different pairs of variational parameters selected from the set \{$\theta_0$, $\theta_1$, $\theta_2$, $\theta_3$\}. On the figures are plotted two convengence trajectories (orange and magenta). }
  \label{Fig_landscape_exp}
\end{figure*}

To validate the accuracy of our convergence, we use the 4-qubit Heisenberg state $S_1$ as an example. The measured potential profiles are compared to theory for 6 different pairs of $\theta$ (4 parameters) in Fig.~\ref{Fig_landscape_exp}\textbf{a-f}. On these figures, we also plot trajectories of two convergence traces, showing how the they reach the minimum in all these dimensions.

\section*{The 4-qubit Ising model}
\begin{figure}
  \centering
  \includegraphics[width=0.45\textwidth]{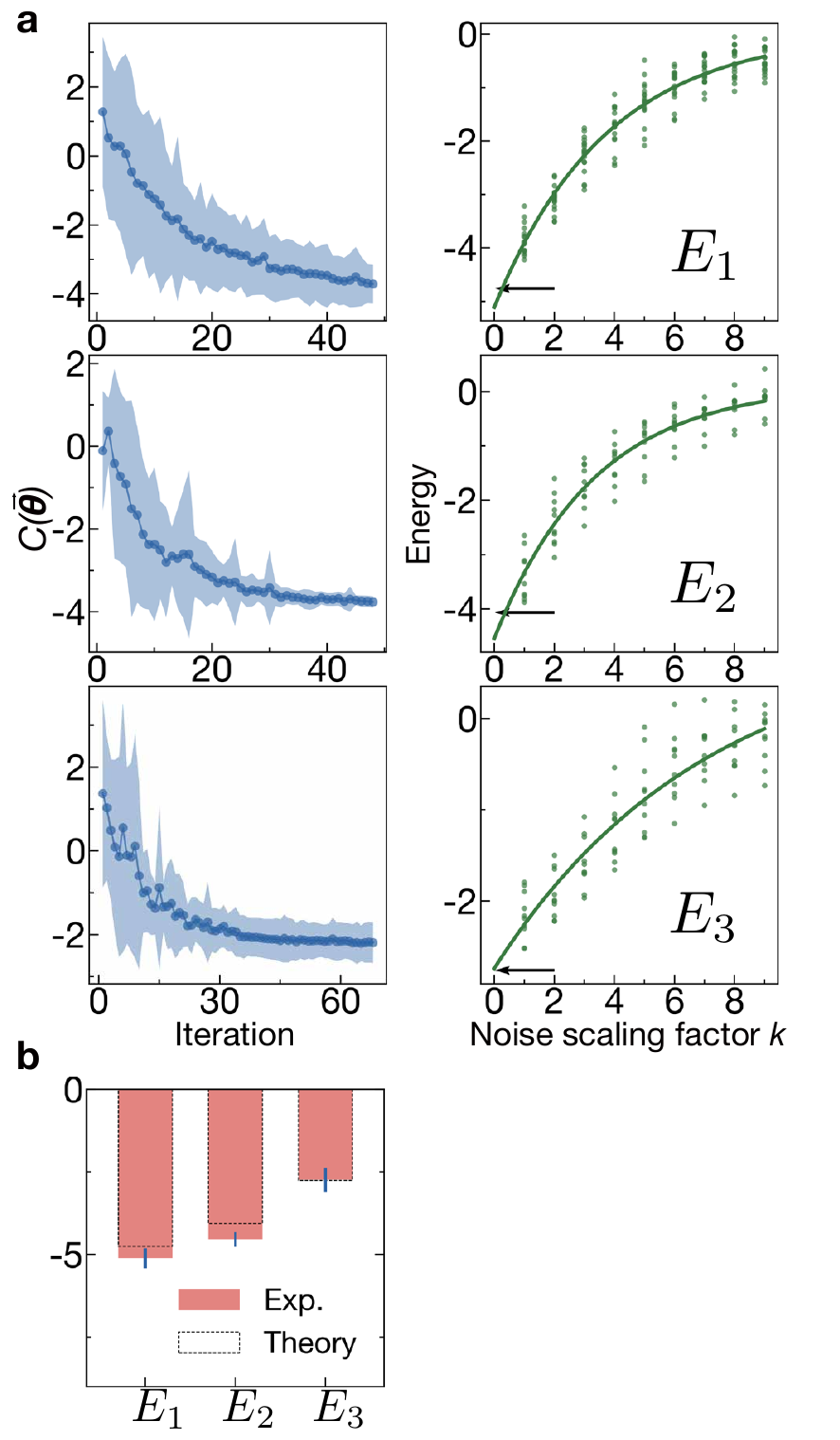}
  \caption[]{\textbf{Low-energy spectroscopy of a 4-qubit Ising spin chain.} 
  \textbf{a,} VQE training (left panels) and error mitigation (right panels) for the three lowest eigenstates for the 4-qubit chain with the transverse Ising interactions.
  \textbf{b,} Comparison of the extrapolated zero-noise energies (red bars) and the theoretical values (dashed box) of the transverse Ising model. The error bars (blue bars) are plus/minus twice the standard deviation from the measured values.}
  \label{Fig_Ising_N4}
\end{figure}

Figure \ref{Fig_Ising_N4} shows the experimental results for the 4-qubit transverse Ising model. With similar training and error mitigation procedures, we obtain the three lowest energy states $E_1$, $E_2$ and $E_3$. The circuits used in the experiment are listed in Table \ref{tab:QinfoN4Ising}. 

\begin{table}[htbp]
\centering
\caption{State properties and VQE information for the 4-qubit and 8-qubit Heisenberg model}
\begin{ruledtabular}
\begin{threeparttable}
\begin{tabular}{c|cccccccc}
4-qubit & $s$ & $s_z$ & $z$& $\langle \hat{H}\rangle$ & $C(\vec{\theta})$ & \# of parameters & Optimizer & \# of $\mathrm{CNOT}^1$  \\ \hline
$\mathcal{S}_1$  & 0 & 0 & 1 & -6.464& $\langle \hat{H}\rangle$  &4&$\mathrm{NM}^2$&18(180) \\ 
$\mathcal{S}_2$ & 0 & 0& 1 & 0.464& $\langle \hat{C_1}\rangle^3$ &12&NM{+}Adam&27(243) \\ 

$\mathcal{T}_1^{(+1)}$   & 1 & 1&-1 & -3.828& $\langle \hat{H}\rangle$ &4&NM&18(207) \\ 
$\mathcal{T}_1^{(0)}$   & 1 & 0& -1 & -3.828& $\langle \hat{H}\rangle$ &4&NM&18(189) \\ 
$\mathcal{T}_1^{(-1)}$   & 1 & -1& -1 & -3.828& $\langle \hat{H}\rangle$ &4&NM&18(207)\\ 

$\mathcal{T}_2^{(+1)}$  & 1 & 1& 1 & -1& $\langle \hat{H}\rangle$ &4&NM&18(207)\\ 
$\mathcal{T}_2^{(0)}$  & 1 & 0& 1 & -1& $\langle \hat{H}\rangle$  &4&NM&18(189)\\ 
$\mathcal{T}_2^{(-1)}$  & 1 & -1& 1 & -1& $\langle \hat{H}\rangle$ &4&NM&18(207)\\ 

$\mathcal{T}_3^{(+1)}$  & 1 & 1& -1 & 1.828& $\langle \hat{H}\rangle$ &4&NM&18(207)\\ 
$\mathcal{T}_3^{(0)}$  & 1 & 0& -1 & 1.828& $\langle \hat{H}\rangle$ &4&NM&18(189)\\ 
$\mathcal{T}_3^{(-1)}$  & 1 & -1& -1 & 1.828& $\langle \hat{H}\rangle$ &4&NM&18(207)\\ 

$\mathcal{Q}^{(+2)}$  & 2 & 2& 1 & 3 & $\langle \hat{H}\rangle$ &2&NM&9(81)\\ 
$\mathcal{Q}^{(+1)}$  & 2 & 1& 1 & 3 & $\langle \hat{C_{2}}\rangle^4$ &8&NM+Adam&18(162)\\ 
$\mathcal{Q}^{(0)}$  & 2 & 0& 1 & 3 & $\langle \hat{C_{2}}\rangle^4$ &8&NM+Adam&18(162)\\
$\mathcal{Q}^{(-1)}$  & 2 & -1& 1 & 3 & $\langle \hat{C_{2}}\rangle^4$ &8&NM+Adam&18(162)\\
$\mathcal{Q}^{(-2)}$  & 2 & -2& 1 & 3 & $\langle \hat{H}\rangle$ &2&NM&9(81)\\
\hline 8-qubit  &&&&&&&
\\
\hline
$\mathcal{S}_1$  & 0 &0& 1 & -13.4997& $\langle \hat{H}\rangle$  & 12&NM&63(603)\\
$\mathcal{T}_1^{(0)}$ & 1 &0& -1 & -11.9289& $\langle \hat{H}\rangle$ & 16&NM&84(792)\\
$\mathcal{T}_2^{(0)}$ & 1 &0& 1 & -10.0149& $\langle \hat{C_{3}}\rangle^5$ & 16&NM&84(792)\\

\end{tabular}

\begin{tablenotes}
    \footnotesize
    \item[1] Number of CNOT gates in one cycle of parameterized circuit. In the brackets is the maximum number of CNOT gates used in error mitigation (up to 9 cycles).
    \item[2] NM = Nelder-Mead algorithm.
    \item[3] $\langle \hat{C_1}\rangle${=}$\omega_1(\langle \hat{H}\rangle_1{+}\beta \langle \hat{S}_{tot}^2\rangle_1){+}\omega_2(\langle \hat{H}\rangle_2{+}\beta \langle \hat{S}_{tot}^2\rangle_2),\omega_1{=}2,\omega_2{=}1,\beta{=}10$
    \item[4] $\langle \hat{C_{2}}\rangle${=}$\langle \hat{H}\rangle+\beta\langle(\hat{S}_{tot}-6)^2\rangle,\beta{=}5$
    \item[5] $\langle \hat{C_{3}}\rangle${=}$\omega_1\langle \hat{H}\rangle_1 + \omega_2\langle \hat{H}\rangle_2,\omega_1{=}2,\omega_2{=}1$
    
\end{tablenotes}

\end{threeparttable}

\label{table:info_heisenberg}
\end{ruledtabular}
\end{table}

\begin{table}[htbp]
\centering
\caption{State properties and VQE information for the 4-qubit and 8-qubit transverse Ising model}
\begin{ruledtabular}
\begin{threeparttable}
\begin{tabular}{c|cccccccc}
4-qubit  & $z$ & $m$ & $\langle \hat{H}\rangle$ & $C(\vec{\theta})$ & \# of parameters & Optimizer & \# of $\mathrm{CNOT}$
\\
\hline
$E_1$  & 1 & 1 & -4.759& $\langle \hat{H}\rangle$ & 8&NM&12(117)&\\
$E_2$ & -1 & -1 & -4.064& $\langle \hat{H}\rangle$ & 8&NM&12(117)&\\
$E_3$ & -1 & 1 & -2.759& $\langle \hat{H}\rangle$ & 8&NM&12(117)& \\
\hline
8-qubit &&&&&&&
\\
\hline
$E_1$  & 1 & 1 & -9.838& $\langle \hat{H}\rangle$ & 24&Adam&42(378)&\\
$E_2$ & -1 & -1 & -9.4689& $\langle \hat{H}\rangle$ & 24&Adam&42(378)& \\
$E_3$ & -1 & 1 & -8.7432& $\langle \hat{H}\rangle$ & 24&Adam&42(378)& \\

\end{tabular}

\end{threeparttable}

\label{table:info_ising}
\end{ruledtabular}
\end{table}

\bibliography{References}